\documentclass[12pt]{elsarticle}
\usepackage{amsmath,amssymb,textcomp,float}
\begin{document}
\title{ Anomalous Brownian Refrigerator }
\author[mymainaddress]{Shubhashis Rana}
\ead{shubho@iopb.res.in}
\author[mymainaddress]{P. S. Pal \corref{mycorrespondingauthor}}
\cortext[mycorrespondingauthor]{Corresponding author}
\ead{priyo@iopb.res.in}
\author[mysecondaryaddress]{Arnab Saha}
\ead{sahaarn@gmail.com}
\author[mymainaddress]{A. M. Jayannavar}
\ead{jayan@iopb.res.in}
\address[mymainaddress]{Institute of Physics, Sachivalaya Marg, Bhubaneswar - 751005, India}
\address[mysecondaryaddress]{Institut f$\ddot{u}$r Theoretische Physik II, Weiche Materie, Heinrich-Heine-Universit$\ddot{a}$t D$\ddot{u}$sseldorf,40225 D$\ddot{u}$sseldorf, Germany}
\begin{abstract}
We present a detailed study  of a Brownian particle driven by  Carnot-type refrigerating protocol operating between two thermal baths. Both the underdamped as well as the overdamped limits are investigated. The particle is in a harmonic potential with time-periodic strength that drives the system cyclically between the baths. Each cycle consists of two isothermal steps at different temperatures and two adiabatic steps connecting them. Besides working as a stochastic refrigerator, it is shown analytically that in the quasistatic regime the system can also act as stochastic heater, depending on the bath temperatures. Interestingly, in non-quasistatic regime,  our system can even work as a stochastic heat engine for certain range of cycle time and bath temperatures. We show that the operation of this engine is not reliable. The fluctuations of stochastic efficiency/coefficient of performance (COP) dominate their mean values. Their distributions show power  law tails, however the exponents are not universal. Our study reveals that microscopic machines are not the microscopic equivalent of the macroscopic machines that we come across in our daily life. We find that there is no one to one correspondence between the performance of our system under engine protocol and its reverse.       
\end{abstract}
\maketitle
\newcommand{\nwc}{\newcommand}
\nwc{\vs}{\vspace}
\nwc{\hs}{\hspace}
\nwc{\la}{\langle}
\nwc{\ra}{\rangle}
\nwc{\lw}{\linewidth}
\nwc{\nn}{\nonumber}

\nwc{\pd}[2]{\frac{\partial #1}{\partial #2}}
\nwc{\zprl}[3]{Phys. Rev. Lett. ~{\bf #1},~#2~(#3)}
\nwc{\zpre}[3]{Phys. Rev. E ~{\bf #1},~#2~(#3)}
\nwc{\zpra}[3]{Phys. Rev. A ~{\bf #1},~#2~(#3)}
\nwc{\zjsm}[3]{J. Stat. Mech. ~{\bf #1},~#2~(#3)}
\nwc{\zepjb}[3]{Eur. Phys. J. B ~{\bf #1},~#2~(#3)}
\nwc{\zrmp}[3]{Rev. Mod. Phys. ~{\bf #1},~#2~(#3)}
\nwc{\zepl}[3]{Europhys. Lett. ~{\bf #1},~#2~(#3)}
\nwc{\zjsp}[3]{J. Stat. Phys. ~{\bf #1},~#2~(#3)}
\nwc{\zptps}[3]{Prog. Theor. Phys. Suppl. ~{\bf #1},~#2~(#3)}
\nwc{\zpt}[3]{Physics Today ~{\bf #1},~#2~(#3)}
\nwc{\zap}[3]{Adv. Phys. ~{\bf #1},~#2~(#3)}
\nwc{\zjpcm}[3]{J. Phys. Condens. Matter ~{\bf #1},~#2~(#3)}
\nwc{\zjpa}[3]{J. Phys. A  ~{\bf #1},~#2~(#3)}
\nwc{\znjp}[3]{New. J Phys. ~{\bf #1},~#2~(#3)}
\nwc{\zapl}[3]{Applied Phys. Lett. ~{\bf #1},~#2~(#3)}
\section{Introduction}

Thermodynamics of micro- and nano-scale systems exhibits distinctly different features from that of large systems due to influence of large thermal fluctuations \cite{rit05}. Typical energy changes are of the order of thermal energy per degree of freedom and consequently  thermodynamics has to be modified at micro-scale. These systems can be theoretically analyzed using stochastic thermodynamics. Exchange of energy between the particle and its surroundings becomes stochastic  and yet one can clearly formulate the notion of work, heat and entropy production for a given microscopic trajectory of the particle \cite{sek98,sei05,dan05,sai07,jop08,arnab09,sek-book,sai11}. Recently obtained exact results (fluctuation theorems\cite{harris07,rit03,jar10,sei12,lah14,lah141}) put constraints on the distributions of the above mentioned stochastic quantities and are valid for systems driven far from equilibrium. These theorems transform thermodynamic inequalities into equalities.
This area has become even more interesting with the development of experimental techniques. Using single-colloidal particle experiments  several new key concepts have been verified. Information to energy conversion and validation of generalized Jarzynski equality \cite{toy10}, Landauer erasure principle \cite{lutz12}, universal features in the energetics of symmetry breaking \cite{rol14} are to name a few. Micron sized heat engines have been experimentally realized by optically controlled
motion of trapped colloidal particle \cite{nphy12,mar14}.

There are several extensive studies on single bath nano-machines e.g., information machines (that can produce work using available information) \cite{dav11,pal14} and molecular motors / thermal ratchets \cite{rei02,mcm}. Molecular motors are omnipresent in cellular as well as tissue level of many living organisms. They are efficient enough to extract energy from a highly fluctuating environment and to convert it into mechanical work for cellular and / or intra-cellular logistics \cite{deba}.

The conversion of energy into mechanical work repeatedly along a cycle working within multiple thermal bath is almost every where in our day-to-day life, spanning huge range of length and time scales. For example, starting from high pressure steam locomotives, to a {\it {drinking bird}} \cite{lor} and even biochemical reaction pathways for cellular respiration mechanism \cite{yoshida,marisa} producing useful energy from nutrients - everywhere energy is transformed into mechanical work. Similarly, by reversing the cycle, we see that in various processes mechanical work is used to transfer heat from a cold source to a warm sink with an objective to cool down the cold source further (refrigerator) or to heat up the warm sink warmer (heat pump).  Though it is very important to study the work-energy (or, vice versa) conversion in all relevant scales, due to lack of experimental techniques for micro or nano world, it is relatively well explored in macro scale.         

Heat engines and refrigerators at nano-scale is a subject of current study \cite{bro06,naka06,mara07,ron08,sch08,eng13,tu13,hol14,luc15}. Detailed theoretical treatment of {\it {Carnot-type}} micro heat engine, involving both quasistatic and non-quasistatic (i.e., finite cycle time) features, have also been documented \cite{sch08,eng13,tu13,hol14,luc15,rana14}. These features reveal the fundamental differences between micro and macro heat engines due to thermal fluctuations, reflected in the distributions of various thermodynamic quantities (e.g., work, heat exchange, efficiency etc.). Unlike macro heat engines, it has been shown that the system can work as a heat engine if the ratio of hot and cold bath temperatures is larger than a critical value \cite{rana14}. Moreover, in non-quasistatic regime, the system works as a heat engine for cycle times larger than a critical value. Both the thresholds depend on the system parameters. Fluctuations in thermodynamic quantities including efficiency of the system 
calculated over a large number of trajectories are significant not only in non-quasistatic regime but also in quasistatic regime, which is in clear contrast to the macro engines. Several trajectories violate typical expectations from the second law of thermodynamics \cite{wang02,sahoo11}. The non self-averaging nature of fluctuations in stochastic efficiency and other quantities requires detailed understanding of full probability distributions as opposed to the average behavior \cite{kumar85,amj91}.
 Large deviation properties of such distributions are recently under theoretical investigations \cite{vernat,vai14,pol15,bro15,verley14,proes,espo}. Research on fluctuation relations for heat engines \cite{sin11,lah12,cam14} are being pursued.  It may also be noted that, for some of the heat
engines studied so far, one may or may not recover Carnot result in the quasistatic regime. However, fluctuation theorems provide a bound on efficiency of an engine valid for any finite time cycle.  Recently, novel theoretical approaches to capture the statistical properties of stochastic efficiency of micro heat engines and mesoscopic thermoelectric engines with broken time-reversal symmetry  are being developed, particularly at long time limit \cite{ver14, jiang15}, claiming universal properties of the large deviation function related to the statistics of stochastic efficiency. To our knowledge, so far, no such study exists for micro-refrigerators. 

\begin{figure}[H]
\begin{center}
\includegraphics[width=9.5cm]{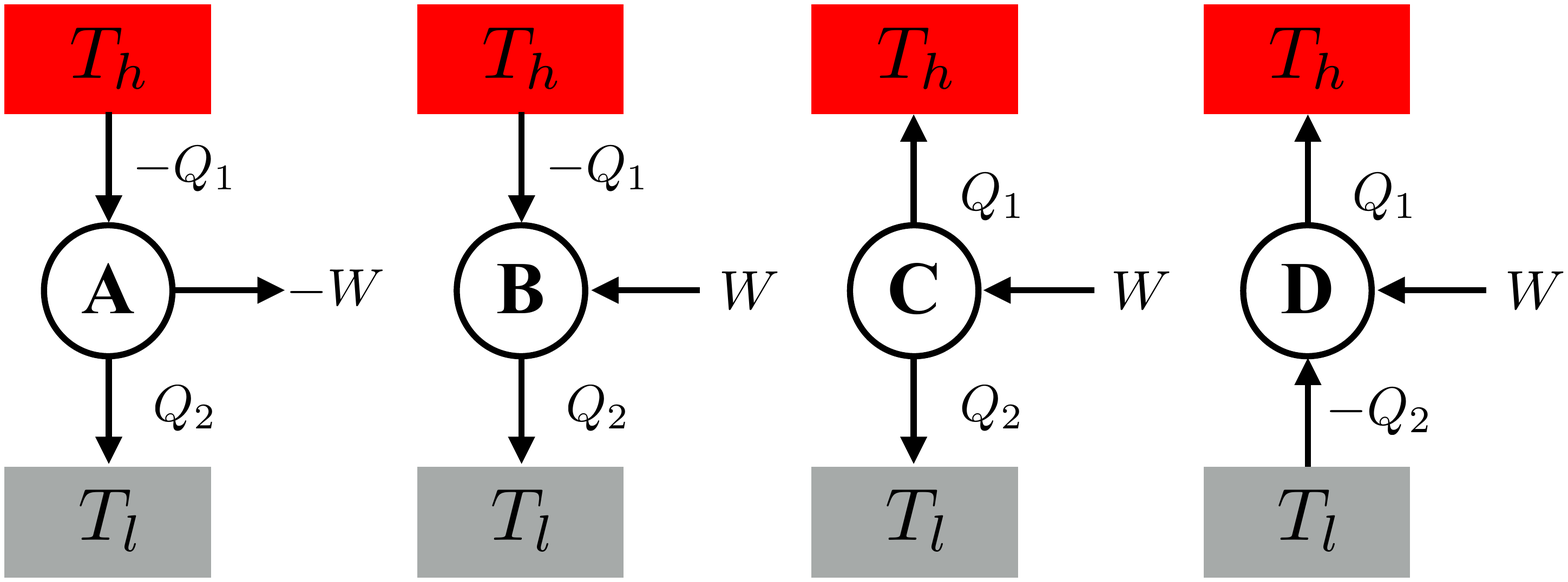}
\caption{(Color online)
Four thermodynamically possible machines working between a hot (with temperature $T_h$) and a cold (with temperature $T_l$) thermal baths: {\bf A}: Takes heat $Q_1$ from hot bath and converts it partially into work $W$ and supplies the rest in the form of heat $Q_2$ to the cold bath (heat engine). {\bf B}: Takes heat $Q_1$ from the hot bath and with the help of work $W$ on it, supplies heat $Q_2$ to the cold bath. {\bf C}: Converts work $W$ on it, to heats $Q_1$ and $Q_2$ that enter into the hot and cold baths respectively.{\bf B} and {\bf C} are called heater of type-II (or, heater-II) and heater of type-I (or, heater-I) respectively.  {\bf D}: Takes heat $Q_2$ from the cold bath and with the help of work $W$ on it, supplies heat $Q_1$ to the hot bath (refrigerator). When work is being done on (by) the system, it is positive (negative). When the system releases heat it is positive and negative otherwise.}
\label{modes}
\end{center}
\end{figure}

In this article we will focus on {\it {Carnot-type}} single-particle refrigerator and its stochastic features. The refrigeration protocol used here is similar to the micro heat engine protocol used in \cite{rana14} but is running backward in time. We believe that our model system is experimentally realizable  using the technique already being used for micro heat  engines. We find new insights into far from equilibrium features of the concerned system. For  example, a major outcome of the present study is the variety of different modes of operation for such systems under the protocol. In Fig. \ref{modes} we describe all the modes of operations which are thermodynamically possible for a system that works cyclically between a hot and a cold heat bath.  Other four possibilities for heat exchanges and work are ruled out due to violation of First and Second laws of thermodynamics.   
Our system, under the refrigeration protocol, exhibits all the modes depending on its cycle time and on the ratio of hot and cold bath temperatures, which can be controlled in experiments \cite{nphy12,mar14}. Throughout the paper, we consider the  heat going from the system to the bath to be positive and that going from the bath to the system to be negative. Similarly, the  work done on the system is taken as  positive and that done by the system is considered negative.    We have also studied 
fluctuation in stochastic efficiency / coefficient of performance (COP) of engine / refrigerator and have shown that  their fluctuations dominate the mean values. Probability distributions have been calculated both in the underdamped and overdamped
 regimes. These distributions exhibit power law tails with non-universal exponents. Reliability of our system operating as a  refrigerator / engine is also reported.     

We first describe the model and the protocol in the next section. In section {\bf 3} we discuss the essentials of stochastic thermodynamics. In section {\bf 4} and {\bf 5}, we explain the results for quasistatic and non-quasistatic behaviour of micro-refrigerators in underdamped as well as in overdamped cases in detail. Finally we conclude by focusing on the major differences between micro and macro machines manifested by the model system. 

\section{The Model}

\begin{figure}[H]
\begin{center}
\includegraphics[width=10.5cm]{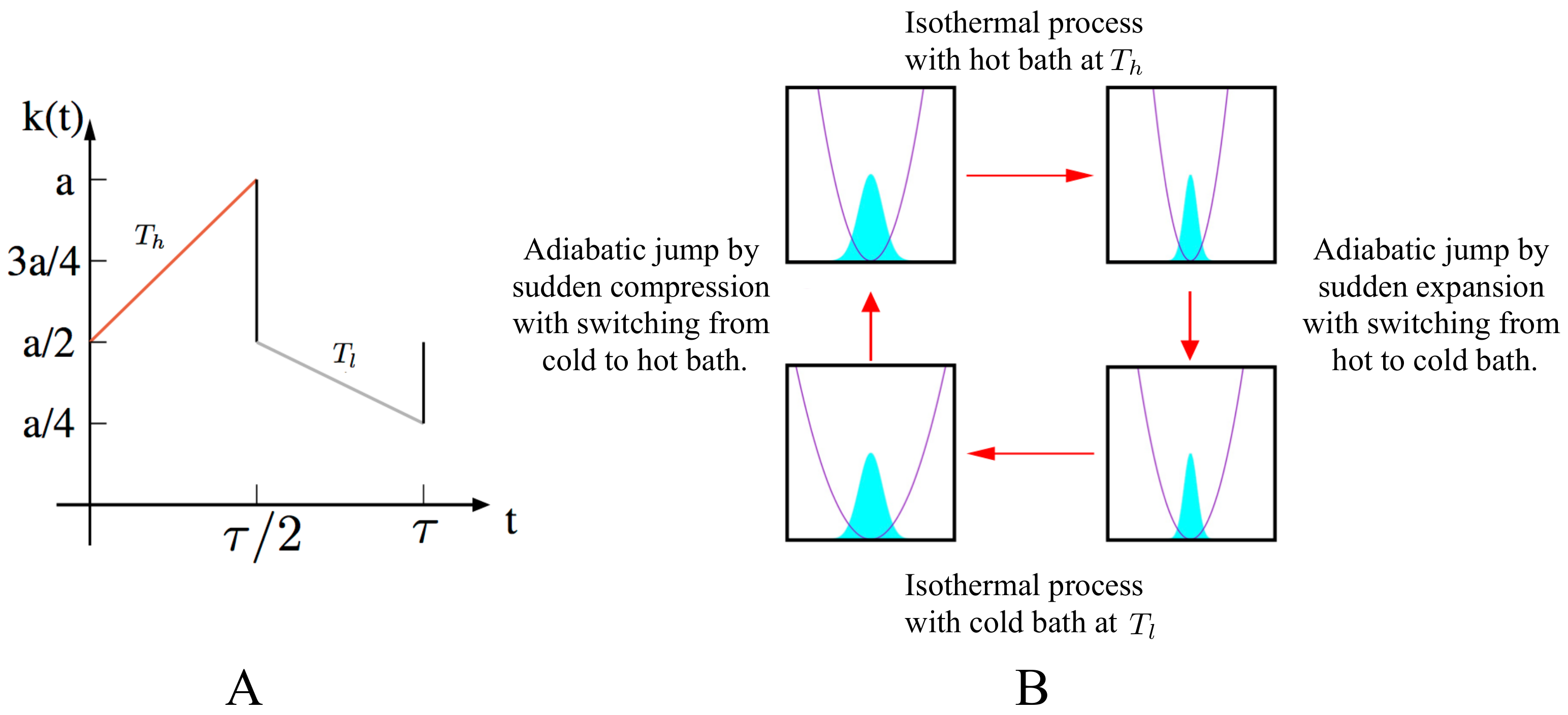}
\caption{(Color online) A. Carnot-type refrigerating protocol: The red and grey lines depict the compression and expansion of the harmonic trap during the isothermal steps with temperature $T_h$ and $T_l$ respectively. Black lines are the jumps of the protocol during the adiabatic steps. B. Schematic representation for a cyclic process of stochastic refrigerator operating between two reservoirs kept at temperatures $T_h$ and $T_l$. The cycle consists of two isothermal steps and two adiabatic steps according to the time varying protocol $k(t)$. The violet line denotes a one dimensional potential $V(x,t)$ and the filled region denotes the corresponding steady state distribution.}
\label{prot}
\end{center}
\end{figure}

We consider a Brownian particle of mass {\it m}, position $x$ and  velocity $v$ moving in a medium with friction coefficient $\gamma$. The particle is trapped  in a harmonic potential whose stiffness constant $k(t)$ is periodically changing with time period $\tau$.  The underdamped equation of motion of the particle in contact with a thermal bath of temperature $T$ is given by \cite{risk-book,coffey-book}
\begin{equation}
m \dot v=-\gamma v-k(t)x+\sqrt{\gamma T}\xi(t),
\label{eom1}
\end{equation}
where fluctuation dissipation relation between noise strength, temperature of the bath and friction coefficient is maintained. Throughout the paper Boltzmann constant $k_B$ is set to unity. In overdamped limit, the equation of motion  reduces to 
\begin{equation}
\gamma\dot x=-k(t)x+\sqrt{\gamma T}\xi(t).
\label{eom2}
\end{equation}
The noises from the bath $\xi$ are  Gaussian distributed with zero mean  and are delta correlated, i.e., $\la \xi(t)\ra=0 $ and $\la \xi(t_1)\xi(t_2)\ra=2\delta(t_1 -t_2)$. The time dependent stiffness  of the trap is given by,  
\begin{eqnarray}
 k(t)&=&a\left(\frac{1}{2}+\frac{t}{\tau}\right) \phantom x\phantom x \phantom x \phantom x 0\leq t < \tau/2  \nonumber\\
 &=&a/2 \phantom x\phantom x \phantom x \phantom x \phantom x\phantom x \phantom x \phantom x  \phantom x\phantom x \phantom x \phantom x t = \tau/2 \nn\\ 
 &=&a\left(\frac{3}{4}-\frac{t}{2\tau}\right) \phantom x\phantom x \phantom x \phantom x \tau/2\leq t < \tau  \nonumber \\
 &=&a/2 \phantom x\phantom x \phantom x \phantom x \phantom x\phantom x \phantom x \phantom x  \phantom x\phantom x \phantom x \phantom x t = \tau .
 \label{protocol1}
\end{eqnarray}
In the first step, the system undergoes an isothermal compression while in contact with the hot bath of temperature $T_h$  for  $0\leq t < \tau/2$. During this step, work is done on the system. At $t=\tau/2$, the stiffness is instantaneously changed from $a$ to $a/2$ (adiabatic expansion) and simultaneously the system is connected to the cold bath at temperature $T_l$, disconnecting it from hot bath. In the third step, isothermal expansion is carried out where the stiffness is changed from $a/2$ to $a/4$ during $ \tau/2\leq t < \tau $. During this step work is extracted from the system. Finally the stiffness is changed by adiabatic compression from $a/4$ to $a/2$. In this step the system is detached from the cold bath and coupled with the hot bath again. A schematic diagram  of the protocol is shown in Fig. \ref{prot}A.
 One should note here that instantaneous jumps are not the only way to implement adiabaticity. There are examples of smooth protocols where one changes the temperature of the bath and the stiffness of the harmonic trap smoothly over time such that at long time limit the phase space volume remains constant, providing no average heat dissipation and consequently establishing the adiabatic steps\cite{mar15}. In this paper we restrict ourselves to this Carnot-type refrigerator protocol. This is the reverse of the protocol that has been studied in \cite{rana14} in context of single particle stochastic heat engine. There are other important protocols, for example Stirling-type protocol used in earlier studies \cite{nphy12}, where the adiabatic steps are absent. The comparative study between stochastic thermodynamics of micro machines driven by different protocols will be focused elsewhere. Our current work concerns stochastic thermodynamics of time periodic steady state (TPSS) of the system. It is well known fact \cite{risk-book} that when a Langevin system is subjected to a time periodic force, after initial transient (duration of which depends on the details of system parameter), the system will settle to a TPSS.  In TPSS, the joint phase space distribution  $P_{ss}(x,v,t) $ is time periodic, i.e., $P_{ss}(x,v,t)= P_{ss}(x,v,t+\tau)$.

\section{Stochastic Energetics of the Single Particle Refrigerator}

Before investigating further, we briefly recapitulate the principle of stochastic thermodynamics. Using Eq. \ref{eom1}, change in  internal energy, $\Delta u$, for an underdamped Langevin system  can be written as 
\begin{eqnarray}
\Delta u &=&\int d\left(\frac{1}{2}mv^2+\frac{1}{2}k(t)x^2\right) \nonumber\\
&=&\int \frac{\partial u}{\partial t}dt +\int (-\gamma v + \sqrt{\gamma T}\xi) v dt   \nonumber\\
& \equiv & w - q 
\label{1stlaw}
\end{eqnarray}     
where $u = \frac{1}{2}mv^2+\frac{1}{2}k(t)x^2$, $w=\int \frac{\partial u}{\partial t}dt $ and $q=-\int (-\gamma v + \sqrt{\gamma T}\xi) v dt$. The integrals in the second step are performed according to Stratonovich's rule. Along a fluctuating trajectory of the particle, identifying the random variables $u$, $w$ and $q$ respectively as total internal energy of the particle, thermodynamic work done on the particle and heat dissipated to the bath, one  obtains first law of stochastic thermodynamics, valid for a single  trajectory (Eq. \ref{1stlaw}). Note that while $w$ and $q$ depend on the entire trajectory of the particle, $\Delta u $ depends only on its initial and final points. Similarly we can also write the first law in case of overdamped dynamics where $u =\frac{1}{2}k(t)x^2$.   
 
For the system concerned here, work done on the particle in isothermal steps for a typical cycle in TPSS is given by,
\begin{eqnarray}
w_{isoth}&=&\int_0^{\tau/2} dt\frac{\partial u}{\partial t}+\int_{\tau/2}^{\tau}dt \frac{\partial u}{\partial t}  \nn\\
&=& \int_0^{\tau/2}dt\frac{1}{2}\left(\dot k x^2\right)_{T=T_h} +  \int_{\tau/2}^{\tau}dt\frac{1}{2}\left(\dot k x^2\right)_{T=T_l} 
\label{isoth}   
\end{eqnarray} 
The first integration in the r.h.s of above expression is defined along the path of isothermal compression and the second one is along isothermal expansion.  By definition, in adiabatic steps $q=0$ and therefore work in adiabatic steps during this cycle in TPSS is  
\begin{eqnarray}
w_{ad}&=&\Delta u  \nonumber\\
&=& \Big[u\Big(\frac{\tau}{2}\Big)-u\Big(\frac{\tau}{2}^{-}\Big)\Big] + \Big[u\left({\tau}\right)-u\left({\tau}^{-}\right)\Big] 
\label{ad}
\end{eqnarray}
The first term of the above expression is the work done on the system along the path of adiabatic expansion while the second one is that along adiabatic compression. Simulating the dynamics of the particle via Eq. \ref{eom1}, we numerically calculate $w_{isoth}$ and $w_{ad}$ to obtain the total work for a cycle as,
\begin{equation}
w=w_{isoth}+w_{ad}.
\label{total}
\end{equation} 
Running the dynamics for a large number of cycles $(N)$ in TPSS, we calculate work, averaged over all cycles,
\begin{equation}
W=\frac{1}{N}\sum_{\text{ all cycles}} w
\label{average}
\end{equation}  
 Using Eq.\ref{1stlaw} we calculate heat transfers in isothermal steps between the particle and the baths ($q$). Heat transfer along isothermal compression is given by 
\begin{equation}
q_{1}= -\int_0^{\tau/2}dt\frac{1}{2}\left(\dot k x^2\right)_{T=T_h}+\Big[u\Big(\frac{\tau}{2}^{-}\Big)-u\Big(0\Big)\Big] 
\label{heat1}
\end{equation} 
and that along isothermal expansion is given by, 
\begin{equation}
q_{2}= -\int_{\tau/2}^{\tau}dt\frac{1}{2}\left(\dot k x^2\right)_{T=T_l}+\Big[u\Big({\tau}^{-}\Big)-u\Big(\frac{\tau}{2}\Big)\Big] .
\label{heat2}
\end{equation}    
Since the heat transfers in adiabatic steps are zero, the total heat transfer in a cycle is
\begin{equation}
q= q_1+q_2.
\end{equation}  
As before, running the cycle repeatedly for $N$ times, we define the average heats,
\begin{equation}
{\mathcal {Q}}_1=\frac{1}{N}\sum_{\text{all cycles}} q_1; {\phantom {xx}} {\mathcal {Q}}_2=\frac{1}{N}\sum_{\text{all cycles}} q_2; {\phantom {xx}} {\mathcal {Q}}=\frac{1}{N}\sum_{\text{all cycles}} q.
\label{heatav}
\end{equation}
These non-quasistatic results should asymptotically match with quasistatic results which can be recovered as $\lim_{\tau \rightarrow \infty}{\mathcal{Q}}_j=Q_j$ with $j=1,2$ and $\lim_{\tau\rightarrow \infty} W=W_{tot}$. Here $Q_j$s are the heat exchanges and $W_{tot}$ is the total work in a cycle in quasistatic limit. Similarly, one can also calculate the change of total energy of the particle in a cycle, $\Delta u$ and its average over $N$ cycles followed by its quasistatic counterpart $U$, taking $\tau \rightarrow \infty$ limit. Therefore, for the system concerned here, all the relevant thermodynamic variables involved in the first law of stochastic energetics and their path-averages, both in non-quasistatic as well as quasistatic regime, can be calculated by simulating the underdamped and overdamped dynamics of the system. Quasistatic behavior of our system serves as an important benchmark for the simulation as it can also be obtained analytically. 

For any machine, quantification of its performance (e.g., efficiency)  is one of the important issues. In case of microscopic engine \cite{rana14}, it has been shown that, in quasistatic regime average efficiency of the micro heat engine depends only on $\frac{T_h}{T_l}$. Similarly, in case of micro refrigerator we will determine average COP in quasistatic regime. We define COP for a cycle as 
\begin{equation}
 \epsilon=\frac{-q_2}{w}
\end{equation}
for a trajectory of the system. It fluctuates randomly from cycle to cycle. For large number of cycles, average stochastic COP is defined as,
\begin{equation}
\la \epsilon \ra=\frac{1}{N}\sum_{\text{all cycles}} \epsilon
\end{equation}  
Due to fluctuations of $q$ and $w$ in both non-quasistatic and quasistatic regime, $\la\epsilon\ra \neq \frac{-\sum q_2}{\sum w}\equiv\bar \epsilon$, where $\bar \epsilon$ is the conventional definition of average COP. For finite time cycle, using fluctuation theorems \cite{sin11,lah12,cam14}, it  can be shown that $\bar \epsilon< \epsilon_c=\frac{T_l}{T_h-T_l}$. Note that this is valid for all cycle times. However, no such bound exists for $\la \epsilon\ra$.  Being equipped with stochastic thermodynamics of our system, in the following sections, we explore quasistatic as well as non-quasistatic behavior of our system.\\ 

\section{Underdamped dynamics}
     
\subsection{Quasistatic Limit}
Now we calculate thermodynamic quantities like, average work, heat exchanges and internal energy changes for different sub steps of a cycle in quasistatic limit.   In this limit, the duration of the protocol is  longer than all other time scales. During isothermal processes,  though the protocol is being changed, the system instantaneously adjusts itself to the equilibrium corresponding to the value of the protocol at that instant. Hence the work done along any isothermal process is the free energy difference between the final and initial state. In the first step, i.e., in isothermal compression, the work done on the system connected to the hot bath will be
\begin{equation}
 W_1=\frac{T_h}{2}\ln\frac{k(t\rightarrow (\tau/2)^- )}{k(t=0)}=\frac{T_h}{2}\ln 2.
 \label{isoth1}
\end{equation}
At $t=(\tau/2)^-$, the system is in equilibrium with hot bath and the corresponding distribution is given by 
\begin{equation}
 P_{\frac{\tau}{2}}(x,v)= N_1\exp\left[-\left(\frac{a x^2}{2}+\frac{m v^2}{2}\right)/T_h\right].
\end{equation}
Here, $N_1=\frac{\sqrt{a m}}{2 \pi T_h}$ is the normalization constant. The second step being instantaneous, there will be no heat dissipation and the work done on the particle is the instantaneous change of its internal energy, given by
\begin{equation}
 W_2=\int\frac{1}{2}\left(\frac{a}{2}-a\right)x^2 P_{\frac{\tau}{2}}(x,v)dxdv=-\frac{T_h}{4}.
\label{ad1}
\end{equation}
Similar to the first step, the work done on the particle in the third step is,
\begin{equation}
 W_3=\frac{T_l}{2}\ln\frac{k(t\rightarrow \tau^- )}{k(t=\tau/2)}=-\frac{T_l}{2}\ln2.
\label{isoth2}
\end{equation}
At $t= \tau^-$, the system will be in equilibrium with the cold bath. Hence, the corresponding distribution will be
\begin{equation}
  P_{\tau}(x,v)= N_2\exp\left[-\left(\frac{a x^2}{8}+\frac{m v^2}{2}\right)/T_l\right],
\end{equation}
with  $N_2=\frac{\sqrt{a m}}{4 \pi T_l}$. Therefore, in the last step, i.e., in adiabatic compression process  the average work done on the particle is
\begin{equation}
 W_4=\int\frac{1}{2}\left(\frac{a}{2}-\frac{a}{4}\right)x^2 P_{\tau}(x,v)dxdv=\frac{T_l}{2}.
\label{ad2}
\end{equation}
The average work done during the full cycle in the quasistatic process is 
\begin{eqnarray}
 W_{tot}&&=W_1+W_2+ W_3 + W_4\nn\\
&&=-\frac{T_h}{4}+\frac{T_l}{2}+\frac{1}{2}(T_h-T_l)\ln2.
\label{tot-w}
\end{eqnarray}
As $T_h>T_l>0$, we find that $W_{tot}$ is always positive for any temperature difference. This implies that, in the quasistatic limit, on an average, work is always done on the system.

To obtain heat absorption from the cold bath ($Q_2$)  first we have to calculate internal energy change along the third step. The average  internal energy at $t=\frac{\tau}{2}^+ $  is $U\left(\frac{\tau}{2}^+\right)=\int\left(\frac{a x^2}{4}+\frac{m v^2}{2}\right)P_{\frac{\tau}{2}}(x,v)dxdv=\frac{3T_h}{4}$. Since the system is in equilibrium with the cold bath at $t=\tau^-$, the average internal energy will be $T_l$. This leads to the change in internal energy in the third step, $\left(T_l-\frac{3T_h}{4}\right) $.
 Using first law (Eq. \ref{1stlaw}) we obtain average heat dissipated to the cold bath, 
\begin{equation}
 Q_2=-\frac{T_l}{2}\ln2 -T_l +\frac{3T_h}{4}.
\end{equation}
Similarly we can obtain  the  average heat, transferred to hot bath, 
\begin{equation}
 Q_1=\frac{T_h}{2}\ln 2 - T_h +\frac{3T_l}{2}.
\end{equation}

\begin{figure}[H]
\begin{center}
\includegraphics[width=6cm]{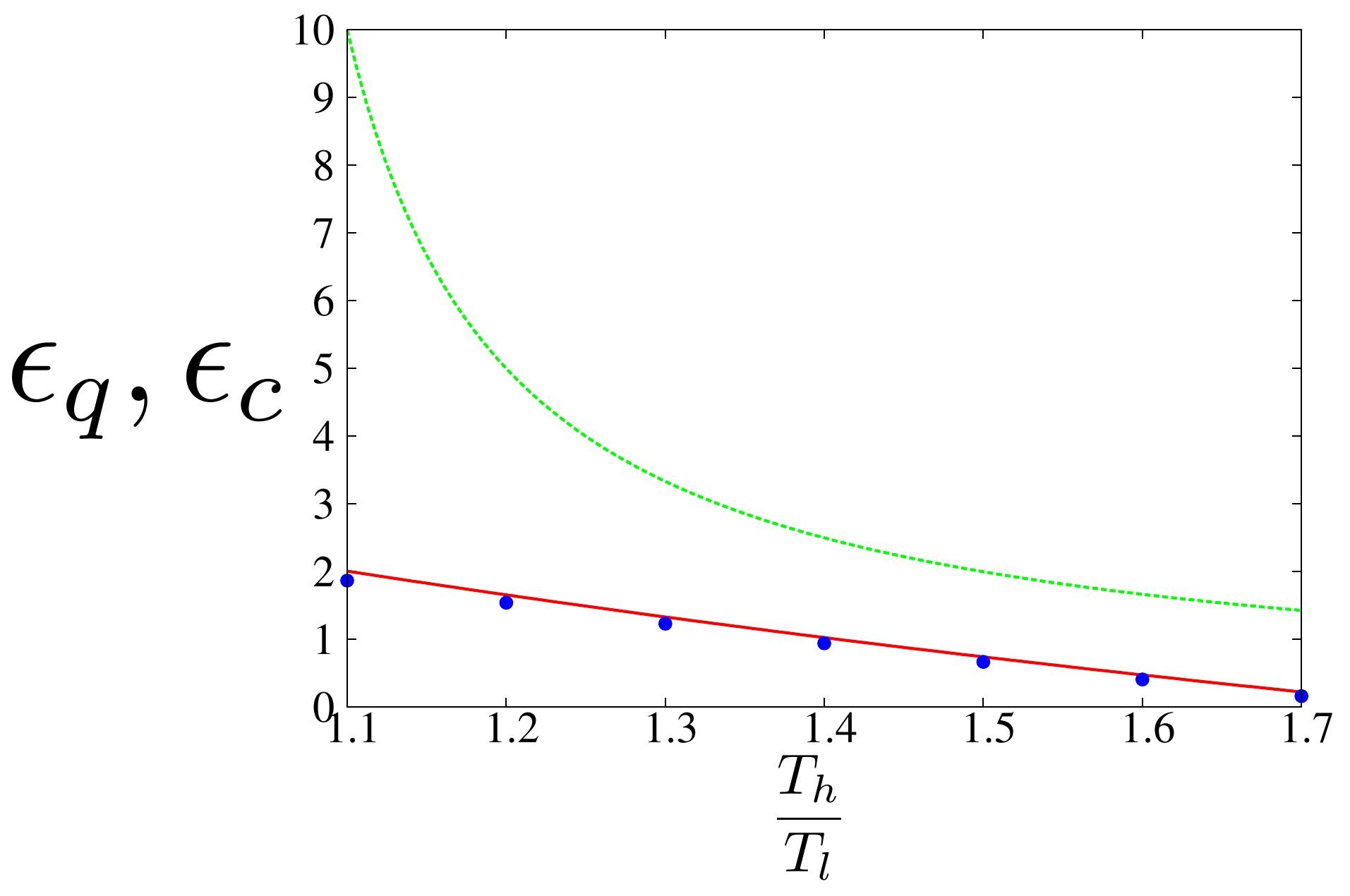}
\caption{ Comparison between quasistatic COP for our model in underdamped limit ($\epsilon_q$, in red line) and the corresponding Carnot bound ($\epsilon_c$, in green line) as a function of $\frac{T_h}{T_l}$. Blue filled circles denotes the values (obtained from numerical simulations) of $\bar{\epsilon}$  at different values of $\frac{T_h}{T_l}$ for large cycle time.}
\label{ud}
\end{center}
\end{figure}

Though $W_{tot} $ is always positive, $Q_1$ and $Q_2$ can take negative as well as positive values depending on the ratio between hot and cold bath temperatures. When $\frac{T_h}{T_l} < 1.80$,  $Q_2$ is negative and is positive elsewhere. When  $\frac{T_h}{T_l} < 2.29$, $Q_1$ is positive and otherwise negative.  The system will act as a refrigerator only when $Q_2<0$ and $Q_1>0$ i.e., $\frac{T_h}{T_l}<1.80$ in the quasistatic limit. For   $1.80< \frac{T_h}{T_l}<2.29$ system will act as a stochastic heater I (Fig. \ref{modes}) where both $Q_1$ and $Q_2$ are positive. When $\frac{T_h}{T_l}>2.29$ heat flows from hot bath to the system $(Q_1< 0)$, but the system releases heat to the cold bath thereby acting as a stochastic heater of type II.  
It is evident that even after using a Carnot-type refrigerating protocol one can obtain single particle refrigerator as well as heaters  depending on $T_h$ and $T_l$ in quasistatic limit. Finally the COP of the system as a refrigerator in quasistatic limit is,
\begin{equation}
 \epsilon_q=\frac{-Q_2}{W_{tot}}=\frac{T_l\ln2 +2T_l -\frac{3T_h}{2}}{-\frac{T_h}{2}+T_l+(T_h-T_l)\ln2}.
 \label{cop}
\end{equation}
It should be noted here that even in the quasistatic limit, COP in our system is much smaller than the  Carnot limit 
($\epsilon_c=\frac{T_l}{T_h-T_l}$) as shown in Fig. \ref{ud}.
We now turn our attention to the results obtained by simulation in non-quasistatic regime.

\subsection{Numerical Results And Discussions}

We evolve the system using discretised Langevin dynamics with time step $dt=0.0002$ in the underdamped as well as overdamped limit [Eq. \ref{eom1} and Eq. \ref{eom2}]. The system is driven by time periodic protocol [Eq. \ref{protocol1}]. We follow Heun's method \cite{heun}. We have set $\gamma=1$ and $m=1$. All the physical quantities are in dimensionless form. Throughout the paper we have fixed $a=5$ and $T_l=0.1$. We make sure that, after the initial transient regime($\sim 10^3$ cycle time), the system settles to a TPSS i.e., $P_{ss}(x,v,t+\tau)=P_{ss}(x,v,t)$. We use the same discretisation to numerically calculate the thermodynamic variables along a trajectory. We  consider at least $10^5$ cycles  of operations and thermodynamical quantities are averaged over all these cycles. 

\begin{figure}[H]
\begin{center}
\includegraphics[width=6cm]{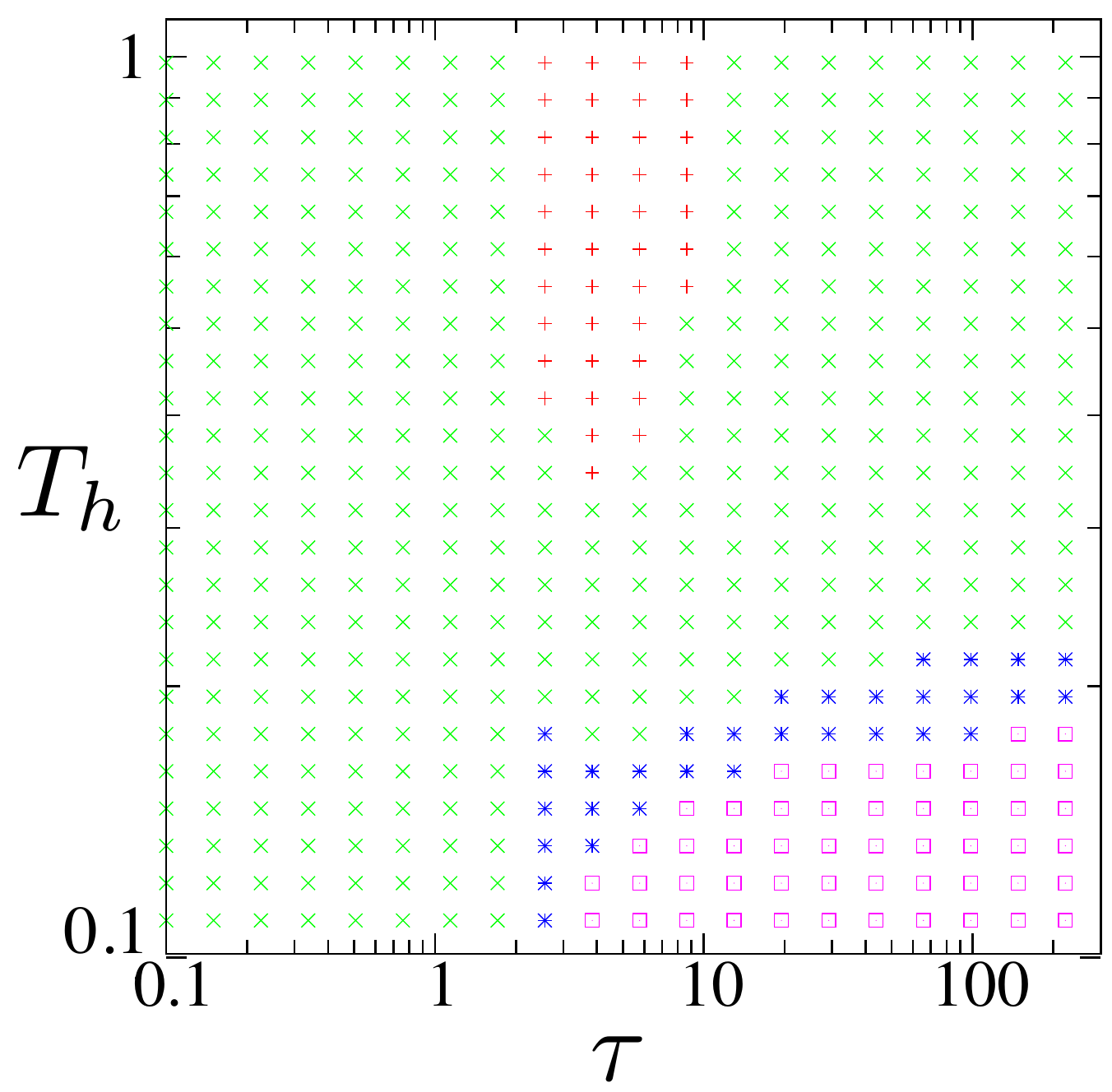}
\caption{Phase diagram for underdamped motion: Different modes of operation of our system following underdamped Langevin dynamics with Carnot refrigerating protocol. Open boxes(pink): refrigerator, asterisk(blue): heater-I, cross(green): heater-II, plus(red): engine. Here all the points in the phase diagram are obtained by averaging over $10^5$ trajectories.}
\label{phase1}
\end{center}
\end{figure}

{\underline{Phase diagram:}} For each $(\tau,T_h)$ pair, we calculate $W$, ${\mathcal{Q}}_1$, and ${\mathcal{Q}}_2$ with $N \sim 10^5$ and a  phase diagram [Fig. \ref{phase1}] for the operational modes of the system is obtained. From the phase diagram it is evident that when $\tau \lesssim 2$, for any $T_h$, ${\mathcal{Q}}_1<0$, ${\mathcal{Q}}_2>0$ and $W>0$. In this mode,  on an average the system heats up the cold bath. Therefore it is  a stochastic heater of type II. 
For large $\tau (\sim 100)$ and high $T_h$, ${\mathcal Q}_1$ and $W$ goes into the system whereas ${\mathcal Q}_2$ comes out of the system to heat up the cold bath, providing us again heater-II. At large $\tau$, but with intermediate $T_h$, we obtain 
heater-I where ${\mathcal Q}_1$ flips its direction of flow to heat up the hot bath keeping everything else the same as before. If we reduce $T_h$ further, we obtain stochastic refrigerator where, even ${\mathcal Q}_2$ flips its direction and flows from 
cold bath to the system while others remain  same as heater-I. Interestingly for $T_h\gtrsim 0.35$ and $2 \lesssim \tau \lesssim 10$,  we observe that on an average, the system can do work, taking heat from the hot bath and releasing it partially to the cold bath. It implies that even if we use a Carnot-type refrigerating protocol, deep in non-quasistatic regime, for a particular range of $\tau$ and $T_h$, our system can behave as a stochastic heat engine.  Therefore, if we consider both non-quasistatic and quasistatic regimes, the system  can act in all thermodynamically possible modes, depending on the values of $(\tau, T_h)$. It should be mentioned here that for large $\tau$, phase boundaries are correctly predicted by the quasistatic limits calculated analytically. The details of the phases and phase boundaries are depicted in Fig. \ref{phase1}. Note that, in the phase diagram crossing any phase boundary once, implies flipping the direction of any one of the ${\mathcal Q}_1$, ${\mathcal Q}_2$ and $W$.  

\begin{figure}[H]
\begin{center}
\includegraphics[width=10cm]{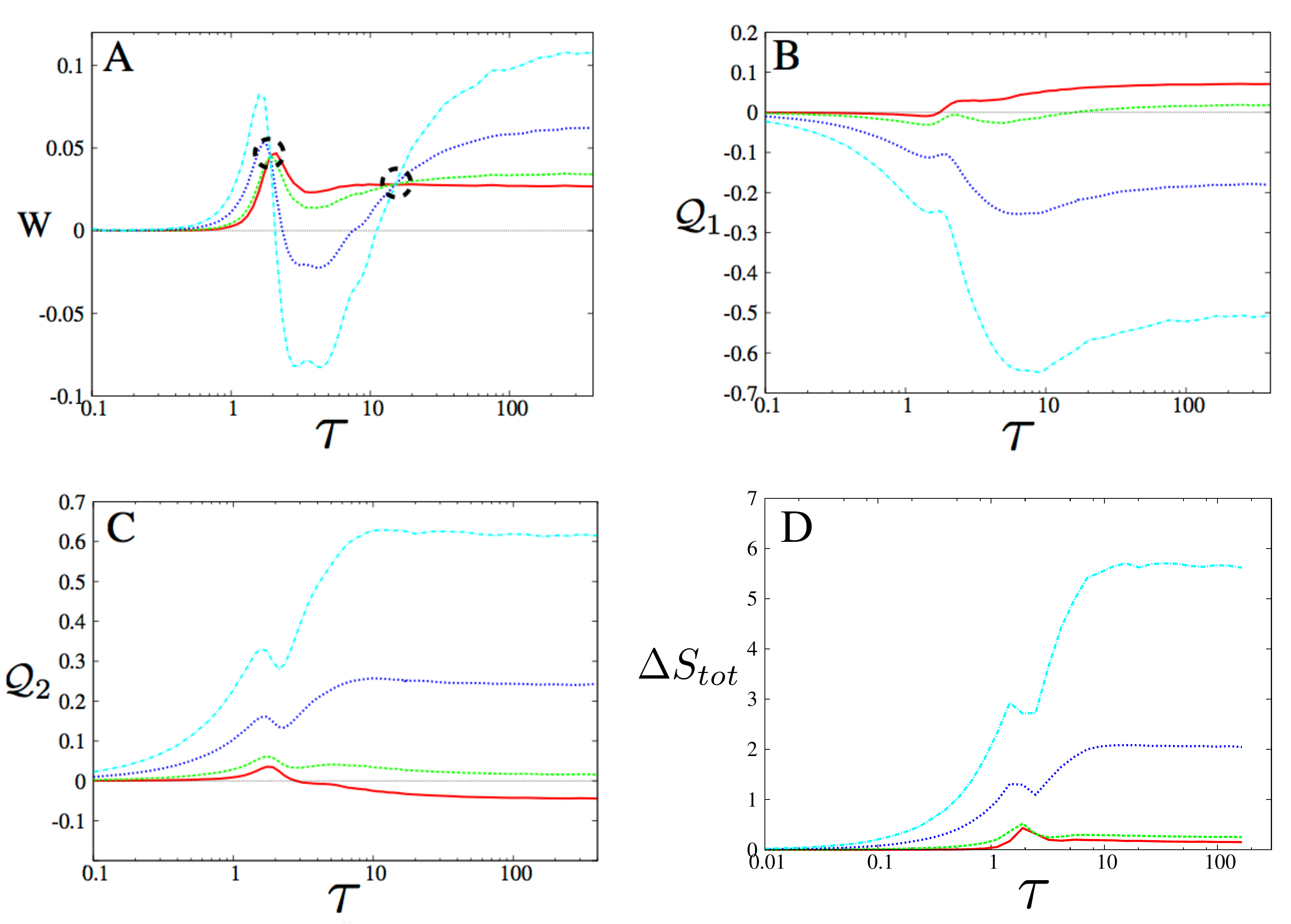}
\caption{ A: $W$ versus $\tau$ for different $T_h$, B:  ${\mathcal{Q}_1}$ versus $\tau$, C:  ${\mathcal{Q}_2}$ versus $\tau$, D:  $\Delta S_{tot}$ versus $\tau$. Red: $T_h=0.12$, Green: $T_h=0.2$: , Blue: $T_h=0.5$, Cyan: $T_h=1.0$). Two crossing points of all the curves in A are marked by  broken black circles in A. At these points $W$ is independent of $T_h$.}
\label{w}
\end{center}
\end{figure}

One can further explore different  modes of operations in Fig. \ref{w} where we have plotted $W$, ${\mathcal{Q}_1}$ and ${\mathcal{Q}_2}$ with respect to cycle time $\tau$, for different $T_h$. If temperature difference between two reservoirs is small, $W$ is always positive and ${\mathcal{Q}_1}$ is negative for small $\tau$ but positive for large $\tau$. The behavior of  ${\mathcal{Q}_2}$ with $\tau$ is  opposite to that of ${\mathcal{Q}_1}$. Thus for small temperature differences, if we vary $\tau$ from small to large values, we obtain heater-II first and then heater-I and finally a refrigerator. For higher temperature differences, with increasing $\tau$,  ${\mathcal{Q}_1}$ becomes negative and ${\mathcal{Q}_2}$ becomes positive. $W$ shows a remarkable non-monotonic (convex) behavior within a range of $\tau$ where it shows a dip that can even be below zero for higher temperature differences. Therefore, in this regime of temperatures, though we obtain heaters with very small and large $\tau$, our system acts as a stochastic heat engine within a particular range of intermediate $\tau$. $W$,  ${\mathcal{Q}_1}$ and ${\mathcal{Q}_2}$ ultimately saturates with $\tau$ to values as found in quasistatic calculation. We note from Fig. \ref{w}A that $W$ becomes independent of $T_h$ for a given $T_l$ at two different $\tau$ within which $W$ is non monotonic.   

\begin{figure}[H]
\begin{center}
\includegraphics[width=12cm,height=4cm]{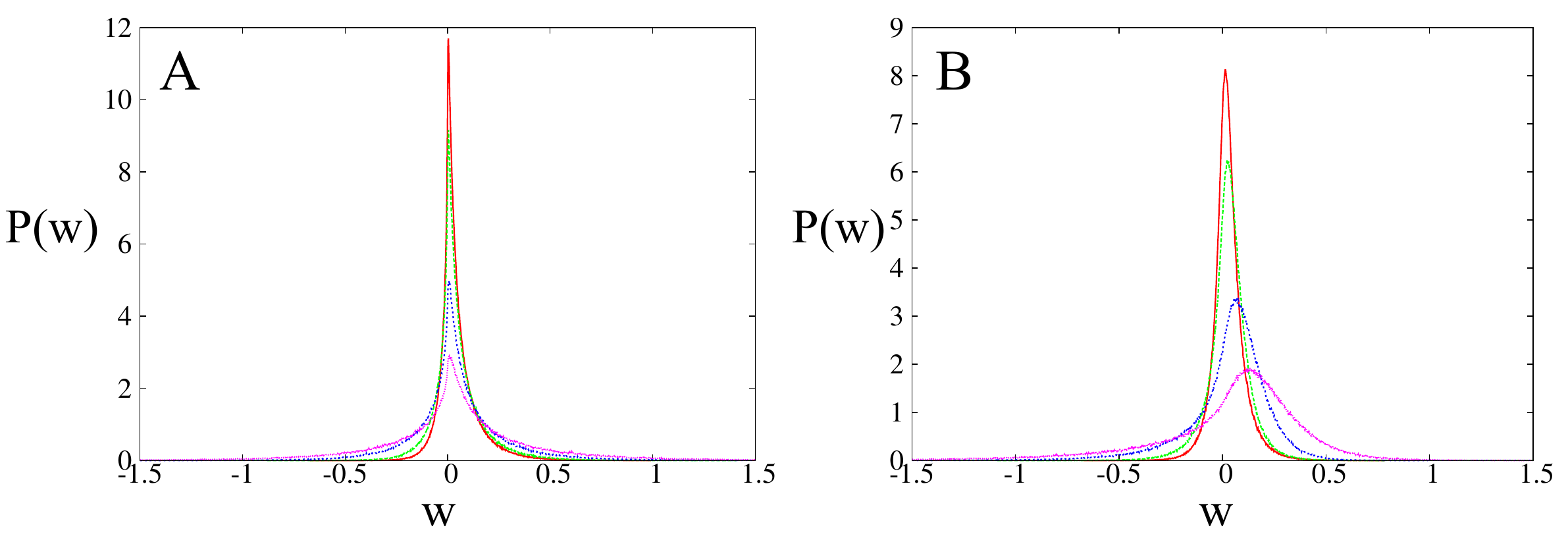}
\caption{Work distributions are plotted for various bath temperatures : $T_h=0.12$ (red), $T_h=0.2$ (green), $T_h=0.5$ (blue), $T_h=1.0$ (cyan). The cycle time for A and B are $\tau=1.916$ (first crossing point) and $\tau=14.98$ (second crossing point) respectively. The distributions are obtained by considering $10^5$ trajectories.}
\label{wdist_un}
\end{center}
\end{figure}

In Fig. \ref{w}D, we have plotted average total entropy production ($\Delta S_{tot}= \frac{\mathcal{Q}_1}{T_h}+\frac{\mathcal{Q}_2}{T_l}$) per period. It shows that initially $\Delta S_{tot}$ increases with $\tau$. Above $\tau=10$, $\Delta S_{tot}$ tends to saturate to quasistatic value.  This sets a characteristic time scale in our system. The characteristic time scale that determine the relaxation time of our system is $\tau\simeq 10$. The entropy production increases monotonically with $T_h$. In Fig. \ref{w}A, we see two special points where the value of $W$ for particular values of $\tau$ are same even though the hot bath temperature $T_h$ is changed. We explore the probability distribution $P(w)$ of work at those two values of $\tau$ for different $T_h$ as shown in Fig. \ref{wdist_un}A and \ref{wdist_un}B. It is clear from the distributions that the fluctuations increases with increasing $T_h$, keeping the mean unchanged. $P(w)$ is asymmetric with positive mean. Finite weight for $w<0$ arises from realisations which doesn't act as refrigerator.

\begin{figure}[H]
\begin{center}
\includegraphics[width=6cm]{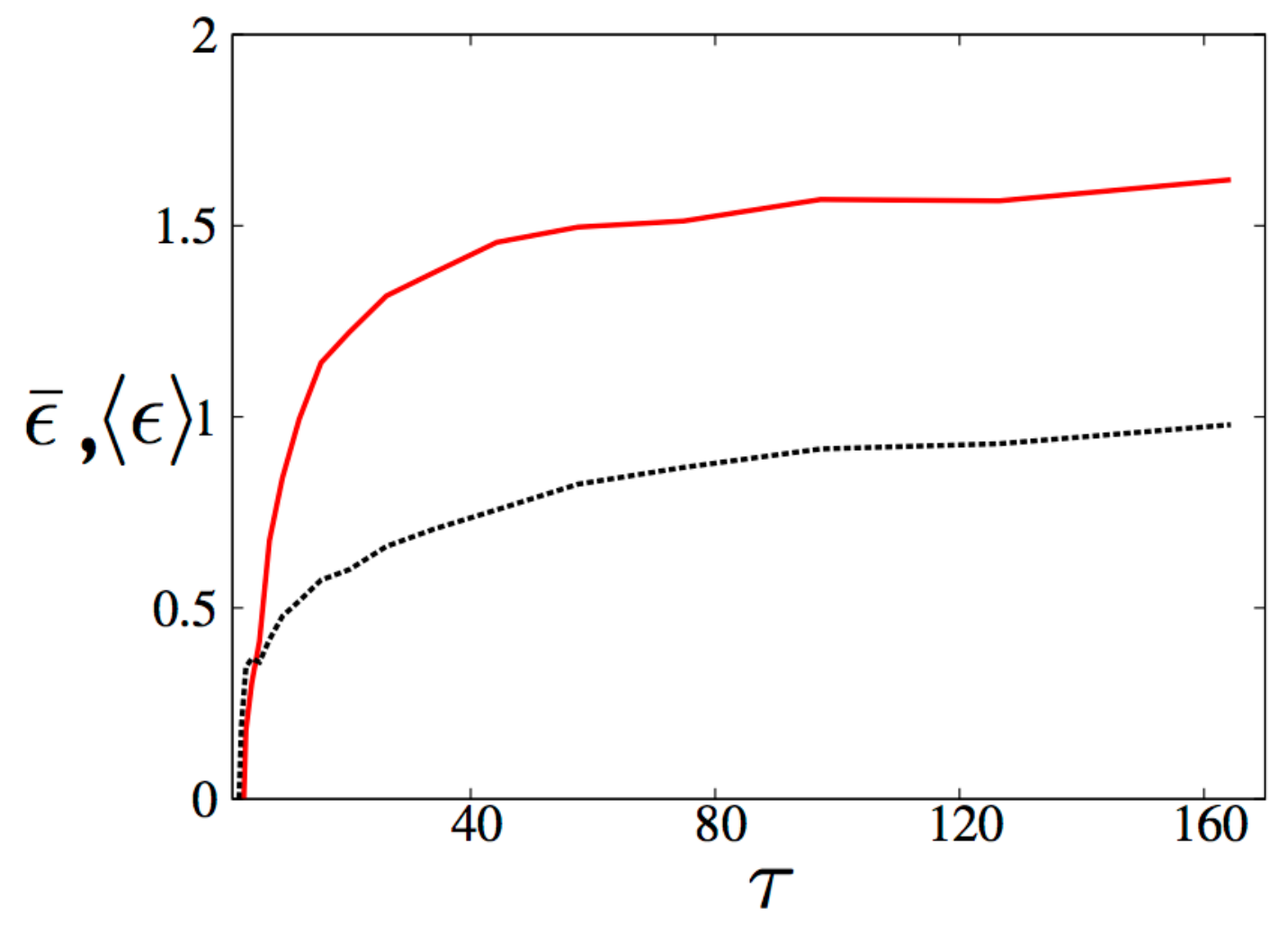}
\caption{Variation of  $\la\epsilon\ra$ (black, broken) and $ \bar\epsilon$ (red, continuous) with $\tau$ at $T_h$=0.12. The average quantities are obtained by considering $10^5$ trajectories.}
\label{coeff}
\end{center}
\end{figure}

{\underline{Refrigerator mode:}} In the refrigerator regime of the phase diagram, most of the trajectories followed by the system maintains the refrigerating condition: $q_1 > 0$, $q_2 < 0$ and $w > 0$. However, there are considerable number of trajectories where the system does not follow this condition. This number reduces with increasing $\tau$ but still remains finite even in quasistatic regime.  Due to fluctuations, the difference between $\la\epsilon\ra$ and $\bar\epsilon$ (Fig. \ref{coeff} A) is prominent 
even at large $\tau$. The saturation values of $\la\epsilon\ra$ and $\bar\epsilon$ (as given in TABLE I) are far below than the Carnot limit ($\epsilon_c=5$ in our case). It should be noted that the saturation value of $\bar\epsilon$ matches with our
quasistatic result.

\begin{figure}[H]
\begin{center}
\includegraphics[width=6cm]{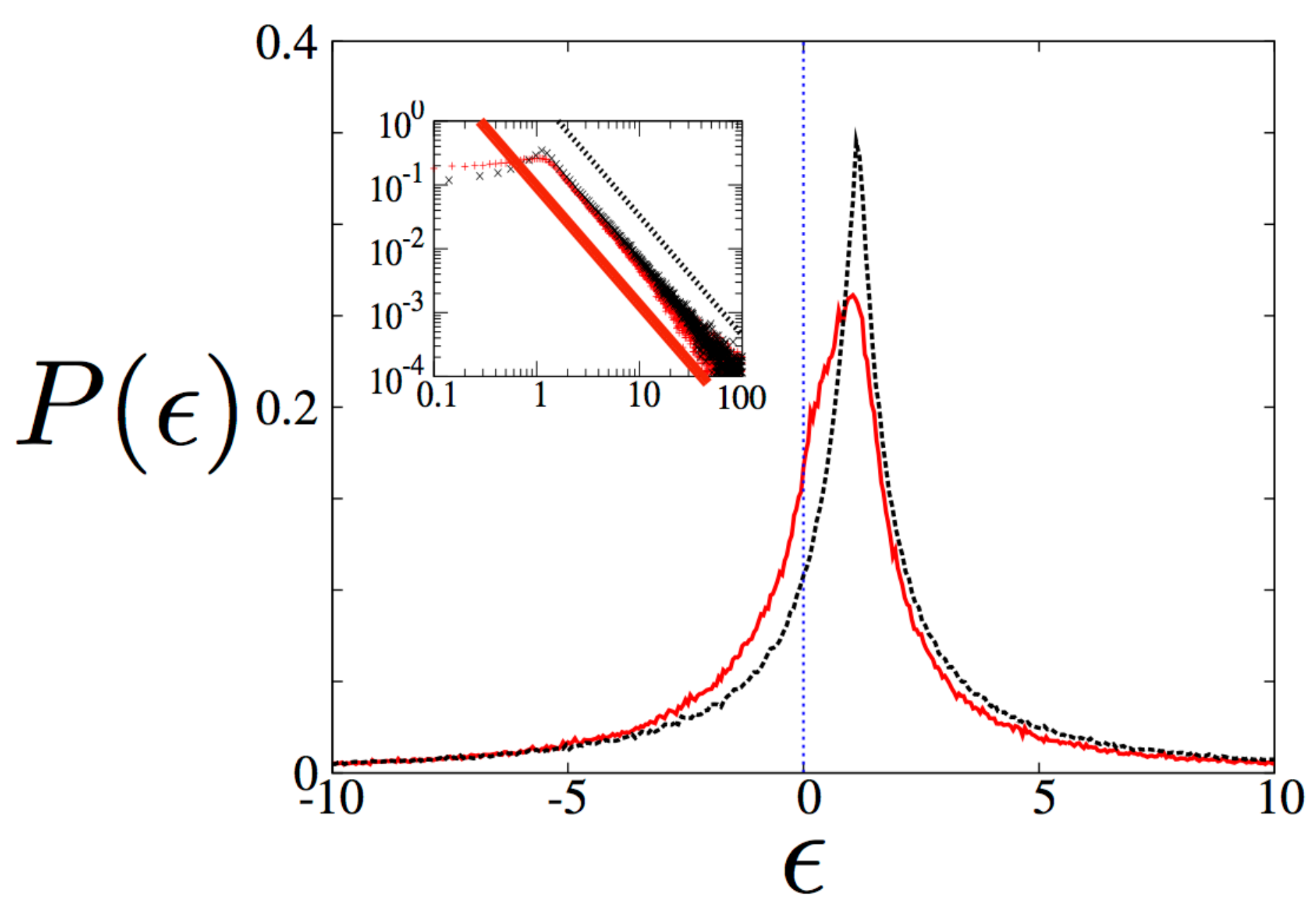}
\caption{Probability distribution of $\epsilon$ at $\tau=10.0$ (red, continuous) and $\tau=100.0$ (black, broken) at $T_h=0.12$.
Inset shows positive tails of the distributions plotted in log scale. They behave as  $\epsilon^{\alpha}$.  For $\tau=10.0$ 
 $\alpha=$-1.897 $\pm$ 0.004 ( slope of the red, continuous line) and for $\tau=100.0$ 
 $\alpha=$ -1.875 $\pm$ 0.004( slope of the black, broken line). }
\label{coeff-dist}
\end{center}
\end{figure}

$P(\epsilon)$, the distribution of stochastic COP, is shown in Fig. \ref{coeff-dist} for two different cycle times, $\tau=10$ and $\tau=100$ (both for $T_h=0.12$).  $\epsilon$ can take values from $-\infty$ to $\infty$. There are considerable number of realizations where $\epsilon$ can be negative or can occur beyond $\epsilon_c$. Moreover, we have noticed (see inset of Fig. \ref{coeff-dist}) that the tail of $P(\epsilon)$ decays as a power law (~$\epsilon^{\alpha}$) for several decades. The exponent $\alpha$ depends on the system and protocol parameters (e.g., $\tau$) and are given in figure captions. So far to our knowledge,  no study exist for the probability distribution and its power law tails for stochastic COP. It is not clear whether this power extends indefinitely for large values of $\epsilon$. However, given our numerical data we can calculate variance which is finite. 

For both distributions, variance $\sigma_{\epsilon} > \la\epsilon\ra$ (see TABLE \ref{table1} for more details). Thus $\la \epsilon\ra$ ceases to be a good physical variable and one has to resort to the study of full probability distribution. From Fig. \ref{coeff-dist} it is also apparent that,  $P(\epsilon)$  becomes sharper with increasing $\tau$.   From the phase diagram we know that, with $\tau \sim 10^2$, our system is closer to the quasistatic limit. Therefore, even in quasistatic regime, fluctuations in $\epsilon$ are significant.

\begin{figure}[H]
\begin{center}
\includegraphics[width=6cm]{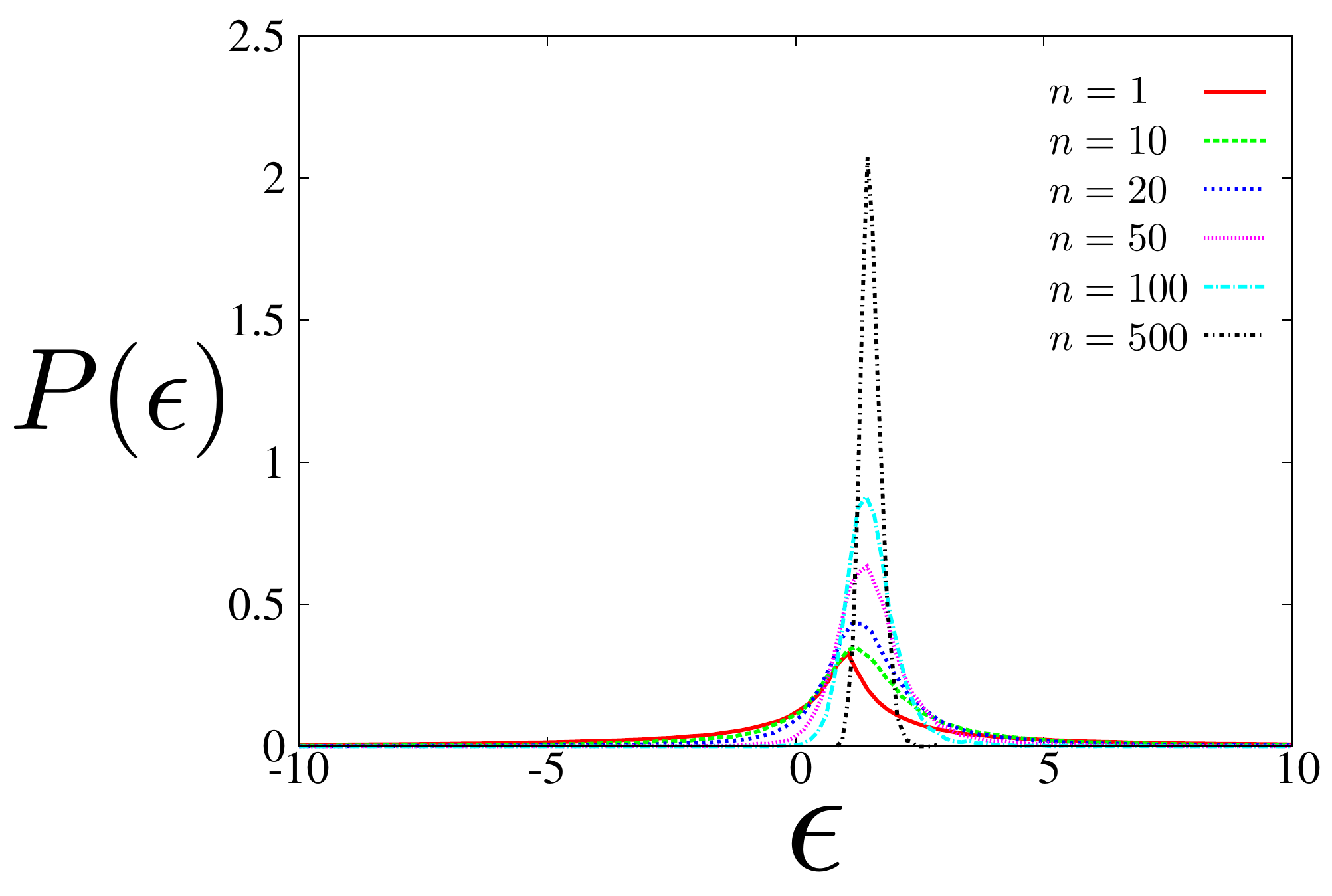}
\caption{Probability distribution of $\epsilon$ for different cycle numbers($n$) at $\tau=100$. }
\label{coeff_dist100_un}
\end{center}
\end{figure}

In Fig. \ref{coeff_dist100_un} distribution of COP is plotted at $\tau=100$ for different numbers of cycle($n$). As $n$ increases, distribution becomes sharper. The tail on the negative side disappears thereby indicating the fact that realisations leading to non-refrigerator mode reduces drastically. Hence our system works as a reliable refrigerator in large $n$ limit. Tail on the positive side also gets suppressed as $n$ increases. The standard deviation decreases as $n$ increases. For $n>50$, standard deviation becomes less than the mean value. This implies that for large $n$, $\langle\epsilon\rangle$ becomes a well defined physical quantity. Average COP does not approach Carnot limit for large $n$.   

\begin{figure}[H]
\begin{center}
\includegraphics[width=12cm]{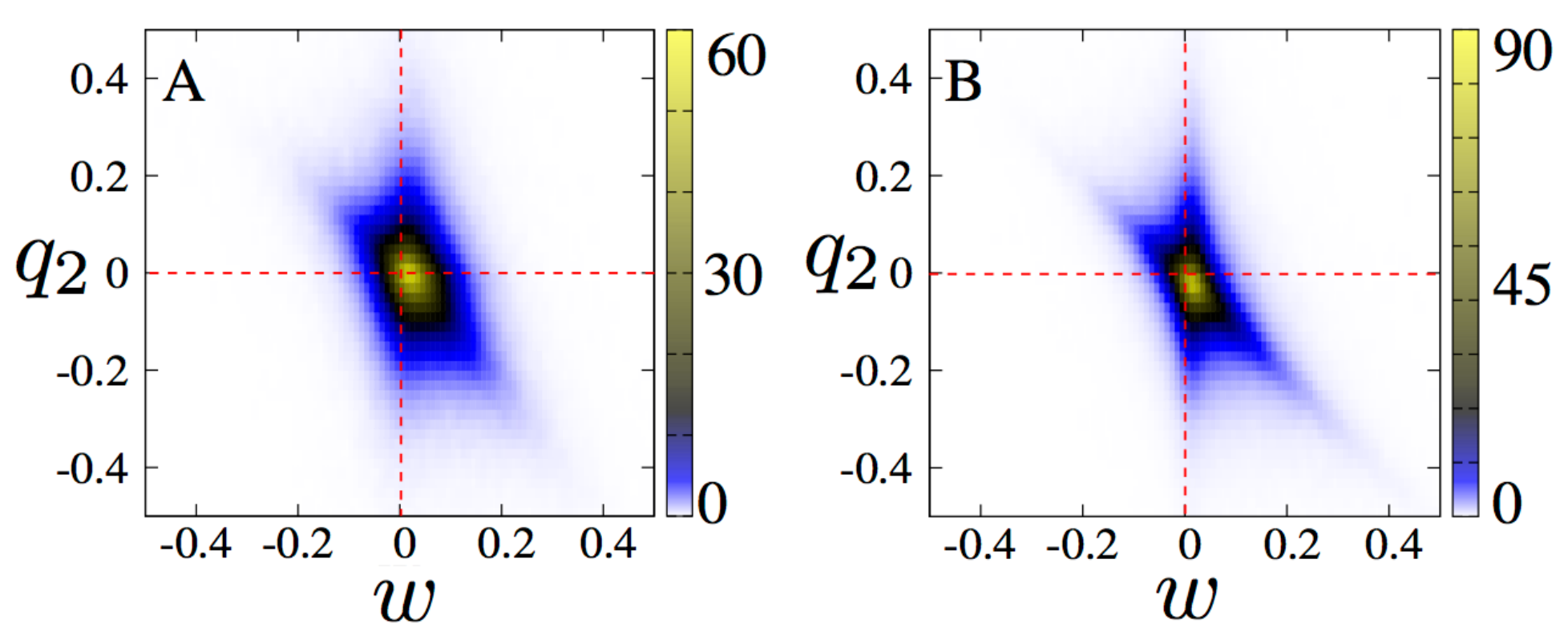}
\caption{Joint distributions:  A: Distribution of $w$ and $q_2$ at $\tau=10.0, T_h=0.12$, B: Distribution of $w$ and $q_2$ at $\tau=100.0, T_h=0.12$.}
\label{ref10}
\end{center}
\end{figure}

 In Fig. \ref{ref10}A and Fig. \ref{ref10}B we have plotted $P(w,q_2)$, the joint probability distributions of $w$ and $q_2$ over $10^5$ trajectories, at $T_h=0.12$ for two different cycle times $\tau=10$ and $100$. For a particular cycle,  $\epsilon$ can be positive   when $q_2$ and  $w$ has opposite signs, but only fourth quadrant of these joint probability distribution represent a refrigerator operation. In Fig. \ref{ref10}A, only  $46.2\%$ trajectories follow the refrigerating condition whereas in Fig. \ref{ref10}B, it is enhanced to $52.1 \%$. Therefore, the system will behave more reliably as a refrigerator if we run it close to quasistatic limit. Table I summarizes  important results.


\begin{table}[ht]
\centering
\caption{}
\begin{tabular}{c c c c c c c c }

 \hline
\hline
$\tau$ & $W$ & ${\mathcal{Q}_1}$ & ${\mathcal{Q}_1}$ & $\bar{\epsilon}$ & $ \la\epsilon\ra$ & $\sigma_{\epsilon}$ & acts as a\\
             &             &             &             &              &                 &              &  refrigerator\\
\hline 

10.0        &   0.0277      &   0.0527     &  -0.0250    &    0.903    &  0.458       &  10.3      & 46.2 \%  \\
100.0       &   0.0271     &    0.0696    &  -0.0423     &    1.561  &   0.929            & 11.7   & 52.1 \%  \\
\hline

\end{tabular}
\label{table1}
\end{table}

\begin{figure}[H]
\begin{center}
\includegraphics[width=6cm]{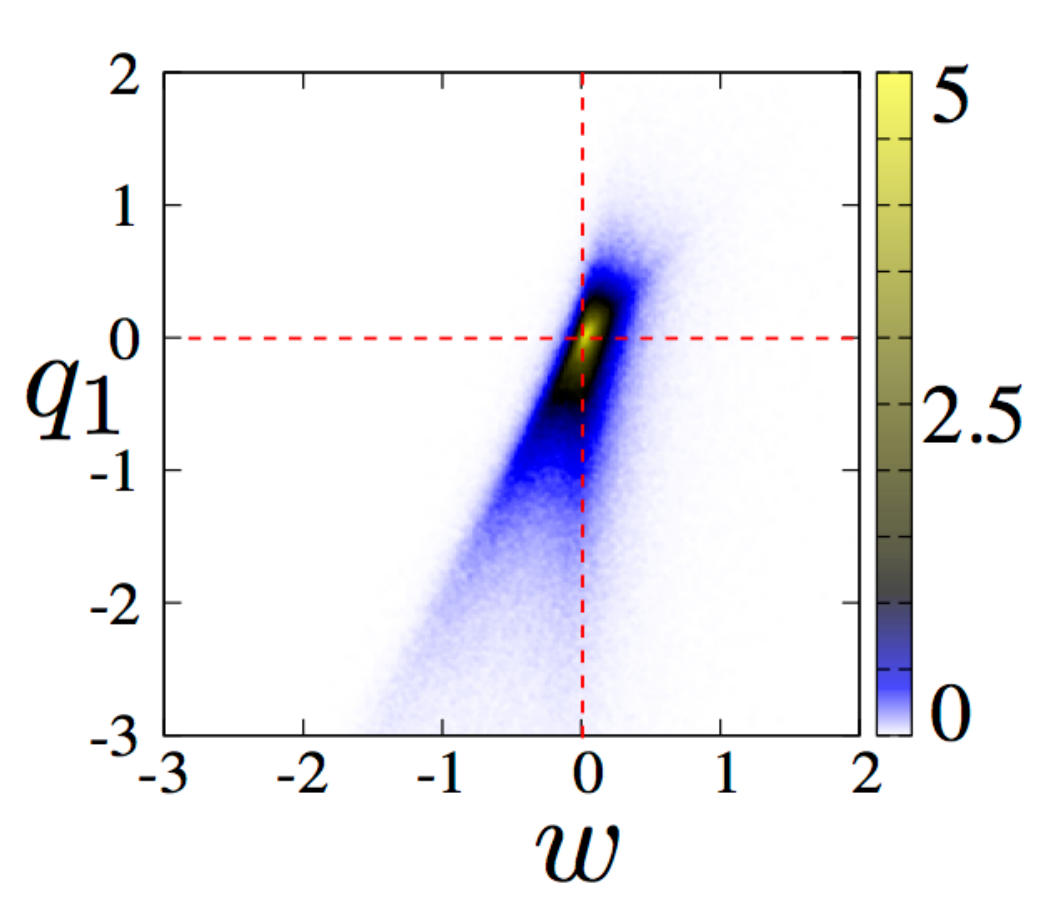}
\caption{Joint distribution of  $w$ and $q_1$ at $\tau=5.0, T_h=1.0$.}
\label{ref11}
\end{center}
\end{figure}

\begin{figure}[H]
\begin{center}
\includegraphics[width=6cm]{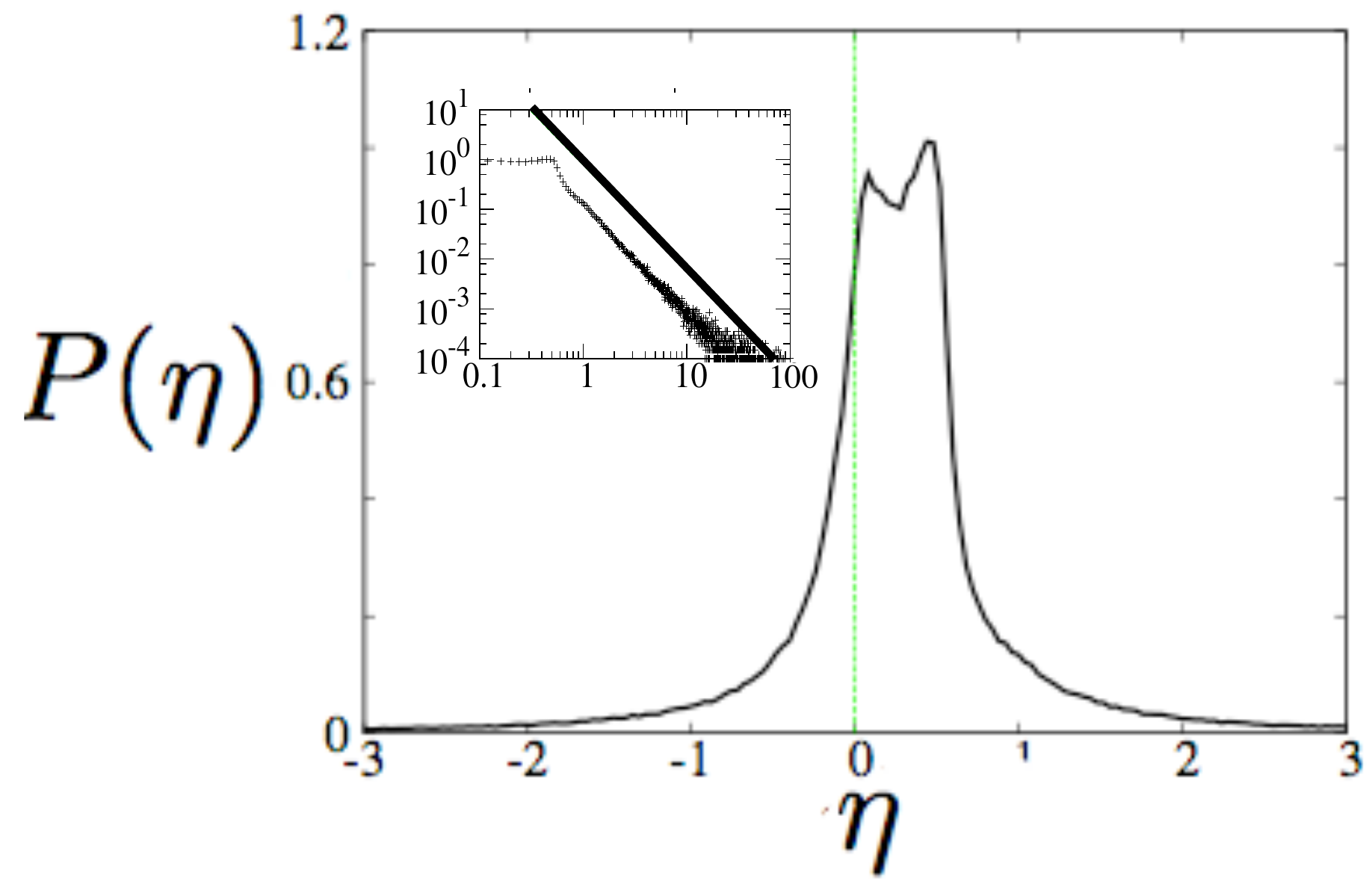}
\caption{Probability distribution of $\eta$ at $\tau=5.0$ at $T_h=1$. Here $\la\eta\ra=0.24$ and $\sigma_{\eta}=3.83$.
Inset: positive tail of the distribution, plotted in logarithmic scale, shows a power law decay with exponent -2.143 $\pm$ 0.010 (slope of the straight 
line).}
\label{effdist}
\end{center}
\end{figure}

{\underline{Engine mode:}} From the phase diagram, it is apparent that the system  acts as an engine only in the non-quasistatic regime where fluctuations are dominant. The joint distribution $P(w,q_1)$ at $(\tau=5.0, T_h=1.0)$, a typical point within engine region of the phase diagram, shows that for only $47.3\%$ of $\sim 10^5$ trajectories, the condition for engine, i.e. $q_1 < 0$, $q_2 > 0$ and $w < 0$,  is followed (Fig. \ref{ref11}).
 The effect of fluctuation is also reflected in  the distribution of stochastic efficiency $\eta$ (Fig. \ref{effdist}). Thus this engine is unreliable. We notice that the efficiency distribution is bi-modal. Both the peaks  as well as $\la\eta\ra$ are well below the Carnot bound $\eta_c=0.9$. Due to strong effect of fluctuations, the standard deviation of efficiency $\sigma_{\eta}$, is much larger than its average $\langle\eta\rangle$. The tail of $P(\eta)$ decays as a power law (~$\eta^{\beta}$) for several decades with $\beta=-2.143\pm0.010$. The exponent is different from that we have obtained in case of $P(\epsilon)$. It suggests that altering $\tau$ and $T_h$ not only implies the change of the thermodynamic mode of operation in $\tau-T_h$ space but it also implies significant difference in the behaviour of probability distributions of their performance index (efficiency / COP).  
 
\section{Overdamped dynamics}

The  difference between the overdamped and underdamped dynamics is  the absence of inertia. The relevant equation of motion for overdamped dynamics of the particle Eq. \ref{eom2} has no acceleration term. This approximation is valid when 
the time steps of the observation are much larger than $m/\gamma$. We now discuss the performance of our system following overdamped dynamics.     

\subsection{Quasistatic Limit}

In the quasistatic limit the calculation of thermodynamic quantities are similar to that of the underdamped case. The main difference is that, in  absence  of inertia, only potential energy contributes to the internal energy. The work done along adiabatic steps will be same as given by Eq. \ref{ad1} and Eq. \ref{ad2}, except for the fact that the probability distribution will now depend only on  position  of the particle not on the velocity. The work done in the isothermal process will be exactly same as in Eq. \ref{isoth1} and Eq. \ref{isoth2}. The expression for total work done on the system in this  process will be the same as that we obtained in the underdamped case (Eq. \ref{tot-w}).

\begin{figure}[H]
\begin{center}
\includegraphics[width=6cm]{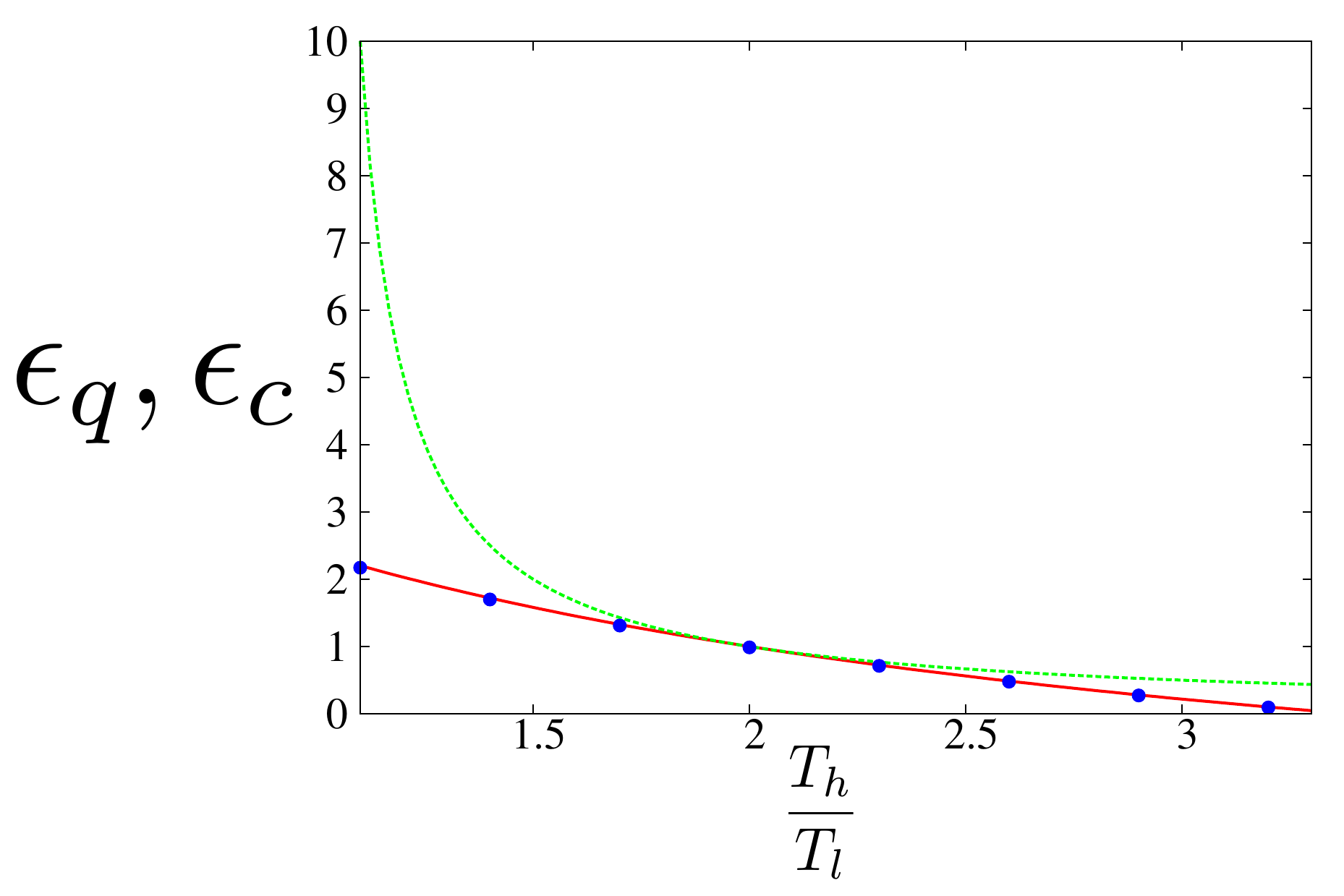}
\caption{ Comparison between quasistatic COP for our model in overdamped limit ($\epsilon_q$, in red line) and the corresponding Carnot bound ($\epsilon_c$, in green line) as a function of $\frac{T_h}{T_l}$. At $\frac{T_h}{T_l}=2$,
 $\epsilon_q$ equals Carnot limit. Blue filled circles denotes the values (obtained from numerical simulations) of $\bar{\epsilon}$  at different values of $\frac{T_h}{T_l}$ for large cycle time.}
\label{od}
\end{center}
\end{figure}

 At $t=\frac{\tau}{2}^+$, average internal energy is $U(\frac{\tau}{2}^+)=\frac{T_h}{4}$. At $t=\tau^-$ system is in equilibrium with the cold bath and corresponding average internal energy is $U(\tau^-)=\frac{T_l}{2}$. The average internal energy change in isothermal expansion process is $\left(\frac{T_l}{2}-\frac{T_h}{4}\right)$. Hence, the average heat that goes to the cold bath in quasistatic limit is given by
\begin{equation}
 Q_2=-\frac{T_l}{2}\ln2 - \frac{T_l}{2} + \frac{T_h}{4}.
\end{equation}
The expression for $Q_1$, the average heat that is exchanged between hot bath and the system,  is
\begin{equation}
 Q_1=\frac{T_h}{2}\ln2 - \frac{T_h}{2} + T_l.
\end{equation}
For overdamped case, the system will act as a refrigerator when  $\frac{T_h}{T_l}<3.39$. For   $3.39<\frac{T_h}{T_l}<6.51$, the system will 
act as stochastic heater-I. Beyond this limit we get stochastic heater-II.  The coefficient of performance for stochastic refrigerator is
 given  by
\begin{equation}
 \epsilon_q=\frac{-Q_2}{W_{tot}}=\frac{T_l\ln2 + T_l - \frac{T_h}{2}}{-\frac{T_h}{2}+T_l+(T_h-T_l)\ln2},
\end{equation}
which is again below the Carnot bound as shown in Fig. \ref{od}. 
We would like to emphasize that only for $\frac{T_h}{T_l}=2$, $\epsilon_q$ equals Carnot limit. This is an interesting observation \cite{sek00}. It is important to note that the adiabatic jumps, being instantaneous, the probability distribution remains unchanged during this process. The system has to relax to new equilibrium along the isotherms after sudden change in the temperature. The relaxation time is negligible in the quasistatic limit i.e. when the cycle time $\tau$ is infinitely large in comparison to the relaxation time. This relaxation leads to an additional heat flow which accounts for the change in the internal energy during the relaxation process. This additional heat flow becomes zero if the system is in equilibrium immediately after the jump. This can be achieved either by changing the jump values of the protocol or by tuning the temperatures. In the overdamped case, we have found that for $\frac{T_h}{T_l}=2$, this additional heat flow is zero and hence the Carnot bound is obtained for our specified protocol. However, in the underdamped limit, one cannot make this additional heat flow equal to zero for any value of adiabatic jump or  $\frac{T_h}{T_l}$ and we do not acheive Carnot bound in this quasistatic limit. We have also found for engine protocol \cite{rana14} that Carnot efficiency $\epsilon_c=1-\frac{T_h}{T_l}$  is achieved only for the value $\frac{T_h}{T_l}=2$ in the overdamped limit. Thus, in the overdamped limit, refrigerators/engines work in reversible mode only for values $\frac{T_h}{T_l}=2$ for our protocol. We will show later that for this particular value, total entropy production over a period vanishes. We now consider the results in non-quasistatic regime obtained in our simulations.

\subsection{Numerical Results and Discussions}
 
\begin{figure}[H]
\begin{center}
\includegraphics[width=6cm]{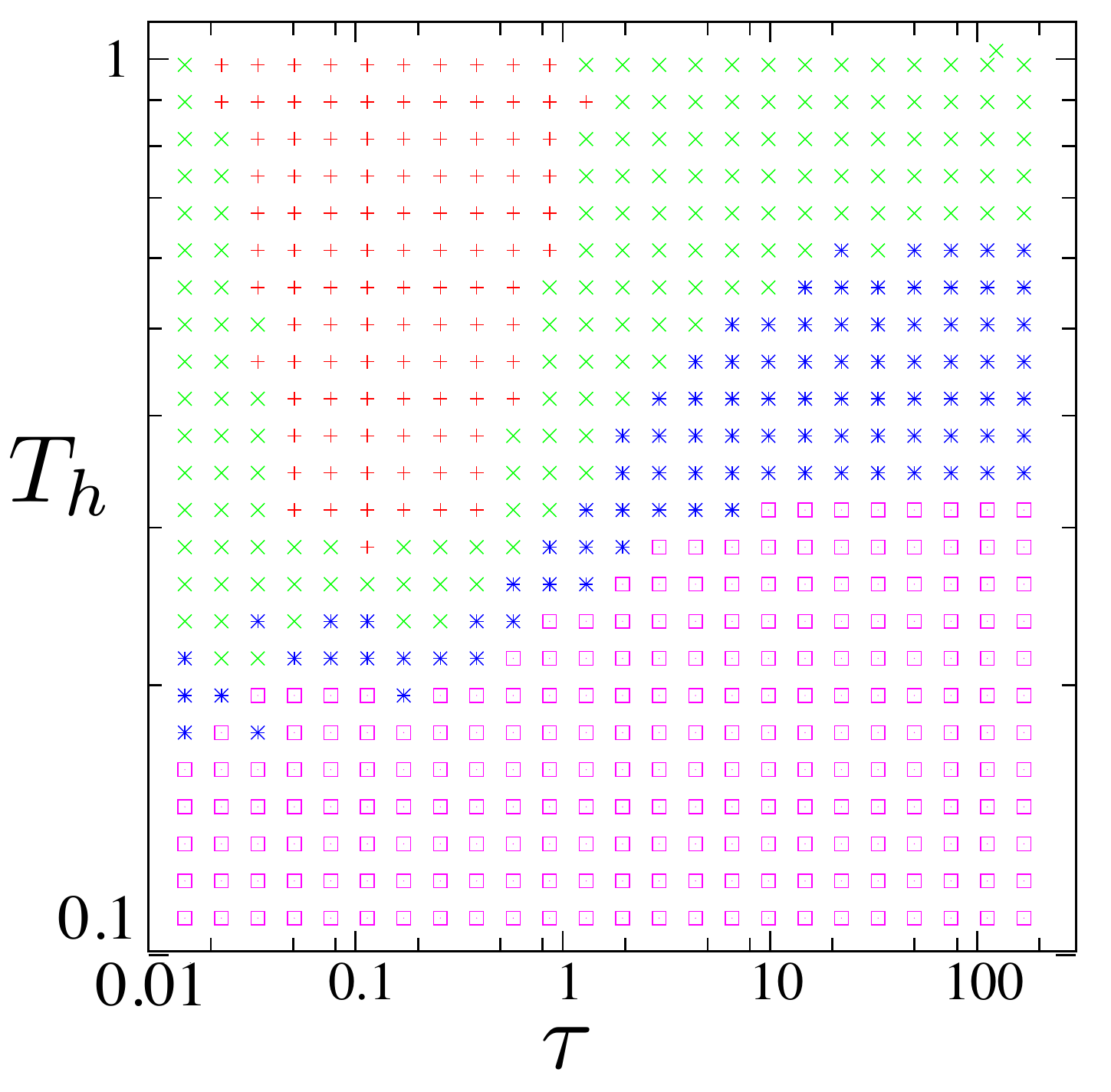}
\caption{Phase diagram for overdamped motion: Different modes of operation of our system following overdamped Langevin
 dynamics with Carnot refrigerating protocol. Open boxes(pink): refrigerator, asterisk(blue): heater-I, cross(green): heater-II, plus(red): engine. The averaging, considered here, is over $10^5$ trajectories.}
\label{ophase}
\end{center}
\end{figure}

We have scanned the parameter space ($\tau-T_h$), keeping $T_l$ fixed at 0.1, to obtain the phase diagram in overdamped regime (Fig. \ref{ophase}). It clearly depicts four different modes of operation of the system, as described in Fig. \ref{modes}.  In contrast to the underdamped case, no critical cycle time is required for the operation in refrigerator mode. The system acts as a refrigerator for  higher temperature differences compared to the earlier case. Therefore, total phase space area of the refrigerator mode has increased in this limit. The phase boundaries in quasistatic limit are consistent with our analytical results. The engine region is clearly visible for  $0.02\lesssim \tau\lesssim 1.0$ and $T_h\gtrsim 0.3$.

\begin{figure}[h]
\begin{center}
\includegraphics[width=10cm]{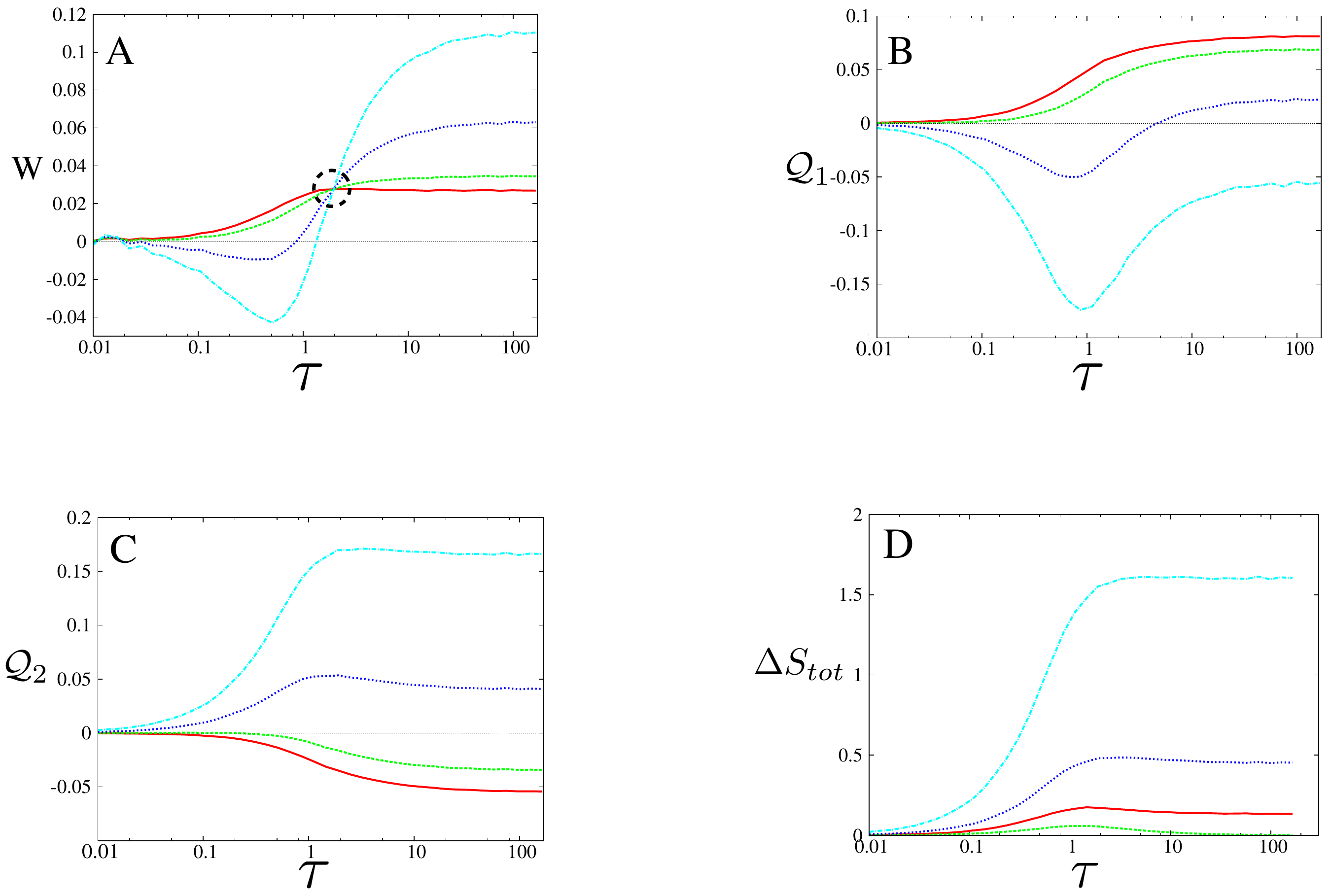}
\caption{ In high friction limit, A: $W$ versus $\tau$ for different $T_h$, B:  ${\mathcal{Q}_1}$ versus $\tau$, C:  ${\mathcal{Q}_2}$ versus $\tau$, D: $\langle \Delta S_{tot}\rangle$ vs $\tau$. Red: $T_h=0.12$, Green: $T_h=0.2$: , Blue: $T_h=0.5$, Cyan: $T_h=1.0$). $W$ becomes independent of $T_h$ at the crossing points of all curves in A, as shown within broken black circle. Here we take the average over $10^5$ trajectories.}
\label{ow}
\end{center}
\end{figure}

\begin{figure}[H]
\begin{center}
\includegraphics[width=6cm]{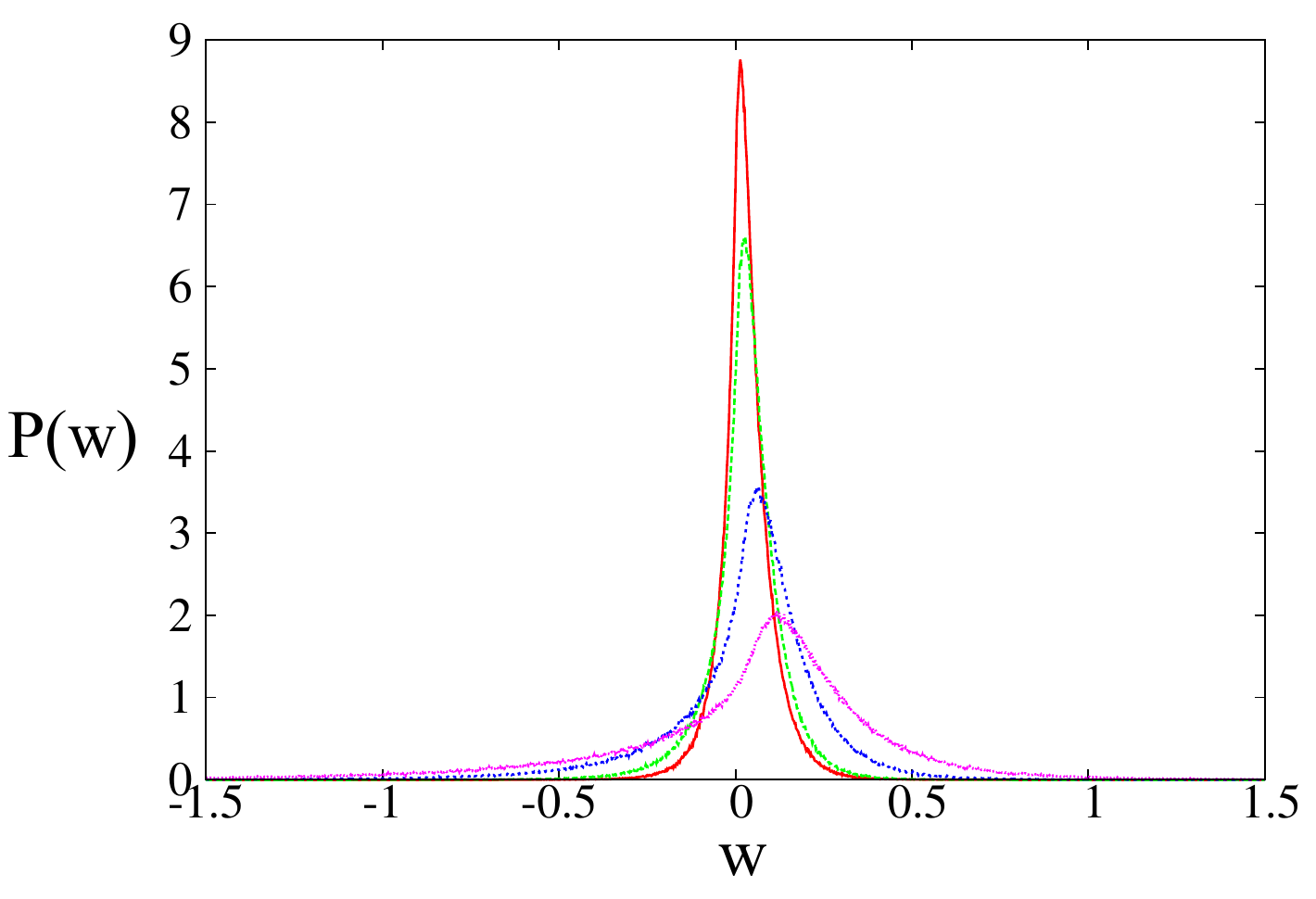}
\caption{Work distributions are plotted at $\tau=1.957$ i.e., at the crossing point of the $W$ vs. $\tau$ plot, for various bath temperatures : $T_h=0.12$ (red), $T_h=0.2$ (green), $T_h=0.5$ (blue), $T_h=1.0$ (cyan).}
\label{wdist_ov}
\end{center}
\end{figure}

To explore  further, in Fig. \ref{ow} we have plotted $W$, ${\mathcal{Q}_1}$ and ${\mathcal{Q}_2}$  as a function of $\tau$ for different $T_h$. Similar to the inertial case, $W$ is positive for all $\tau$ at low $T_h$(=0.12,0.2). However at higher values of $T_h$(=0.5, 1.0), non-monotonicity of $W$  with respect to $\tau$ becomes apparent. For smaller $\tau$ it is negative, showing a dip around $\tau\sim 0.8$. After the dip, $W$ increases and eventually saturates for large $\tau$ at a positive value, predicted by quasistatic result. Interestingly, in high friction limit, the range of $\tau$ where $W$ is non-monotonic, starts from $\tau \simeq 0$ whereas in the underdamped case, the range was bounded between two nonzero cycle times. From Fig. \ref{ow} it is apparent that behavior of $W$, ${\mathcal{Q}_1}$ and ${\mathcal{Q}_2}$ with $\tau$, are all consistent with the phase diagram for different modes of our system in the high friction limit.
\\
In Fig. \ref{ow}D we have plotted $\Delta S_{tot}= \frac{\mathcal{Q}_1}{T_h}+\frac{\mathcal{Q}_2}{T_l}$ as a function of $\tau$. The characteristic time scale that determine the relaxation time of our system in overdamped limit is $\tau\simeq 5 $. For $\tau\geq5$, $\Delta S_{tot}$ tends to saturate to quasistatic value. Interestingly, for $\frac{T_h}{T_L}=2$, $\Delta S_{tot}$ approaches to zero in the quasistatic limit. It implies, for this special case, refrigerator works in a reversible mode and the corresponding COP equals Carnot value as discussed earlier. For all other values of $\frac{T_h}{T_L}$, our system works in an irreversible mode with finite value of $\Delta S_{tot}$ even in quasistatic limit therby preventing it to reach Carnot value. In Fig. \ref{ow}A, we see that at $\tau= 1.957$, $W$ does not change with $T_h$. In Fig. \ref{wdist_ov} we have plotted $P(w)$ for different $T_h$ to show that the mean is constant but the width increases with $T_h$. $P(w)$ is asymmetric with positive mean. Finite weight for $w<0$ comes from realisations which doesn't act as refrigerator.

\begin{figure}[H]
\begin{center}
\includegraphics[width=6cm]{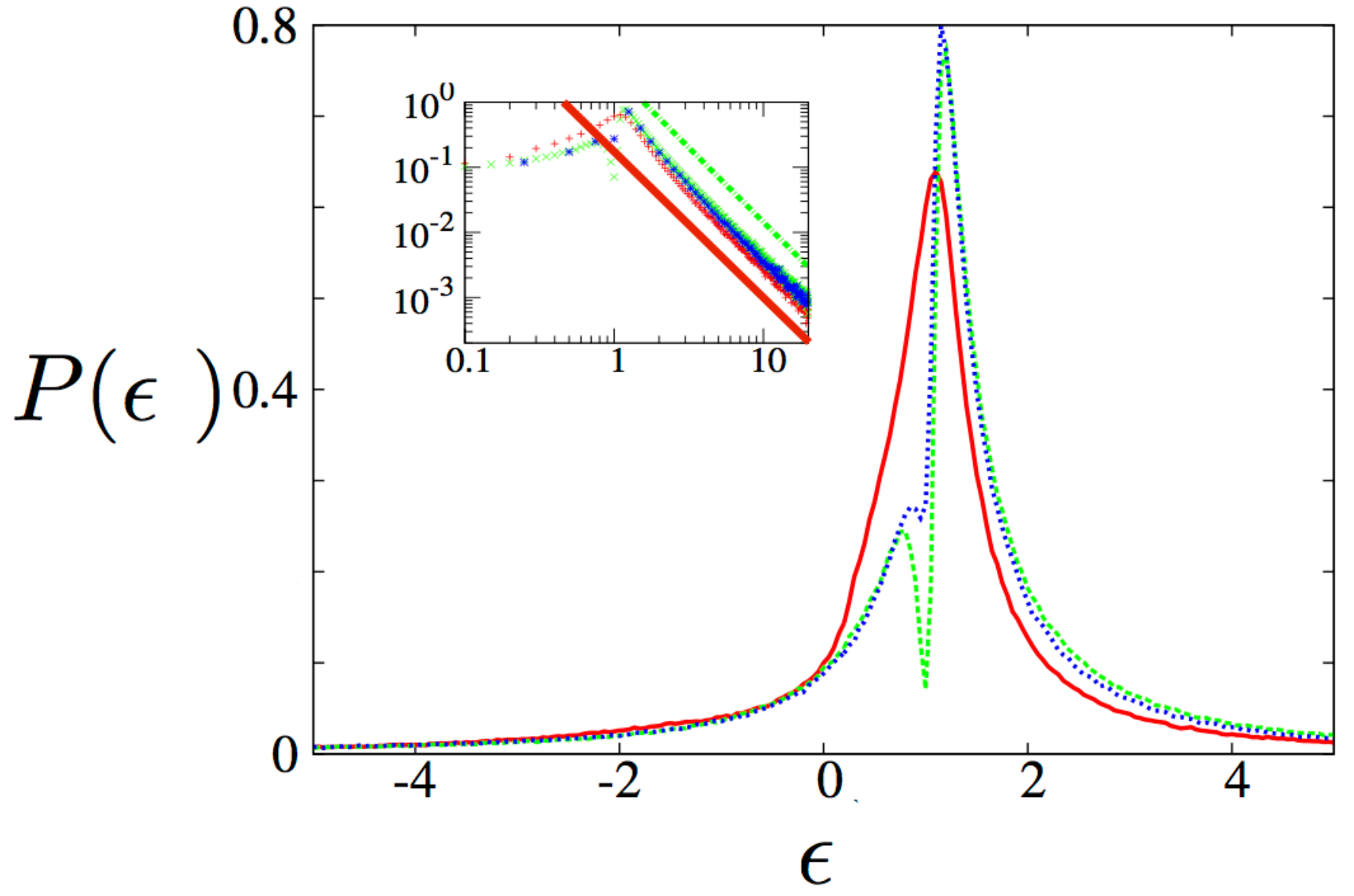}
\caption{Probability distribution of $\epsilon$ at $\tau=10.0$ (red, continuous), $\tau=50.0$ (blue, broken) and $\tau=100.0$ (green, broken) at $T_h=0.12$. Inset:positive tails of the distributions plotted in log scale. Slopes of the straight lines indicate the exponents. For $\tau=10.0$, 50.0 and 100.0, $\alpha=$-2.255 $\pm$ 0.010 (red, continuous line), -2.275 $\pm$ 0.010, -2.288 $\pm$ 0.008  (green, broken line)  respectively.}
\label{ocoeff-dist}
\end{center}
\end{figure}

In Fig. \ref{ocoeff-dist} we have plotted the distribution of $\epsilon$, $P(\epsilon)$, for $\tau=10,50$ and 100. Close to quasistatic limit ($\tau=100$) we observe a dip in $P(\epsilon)$ at $\epsilon=1$ and the distribution becomes bi-modal. The fluctuations prevail even at large $\tau$, thereby making $\epsilon$ a non-self-averaging quantity. For large $\epsilon$, $P(\epsilon)$ shows power law tails with exponents given in figure caption. Notably exponents are lesser than $-2$, indicating finite variance. The exponent values are different from that obtained in the underdamped case. This suggests that the exponents are not universal as our system works in an irreversible mode.  Mean and standard deviations of different stochastic thermodynamical variables for two different values of $\tau$ at $T_h=0.12$  are given in Table 2.

\begin{table}[ht]
\centering
\caption{}
\begin{tabular}{c c c c c c c c }
 \hline
\hline
 $\tau$ & $ W $ & ${\mathcal{Q}}_1$ & ${\mathcal{Q}}_2$ & $\bar{\epsilon}$ & $\la \epsilon \ra $ & $\sigma_{\epsilon}$ & acts as a \\
             &             &             &             &              &                 &              &  Refrigerator\\
\hline 
10.0        &    0.0272    &   0.0766     &   -0.0494      &   1.830   &   0.916   &  6.863   &    65.6 \%  \\
100.0       &  0.0270       &   0.0812     &   -0.0541     &   2.004  &    1.348   &    7.80        & 67.1 \%  \\
\hline
\end{tabular}
\end{table}

\begin{figure}[H]
\begin{center}
\includegraphics[width=6cm]{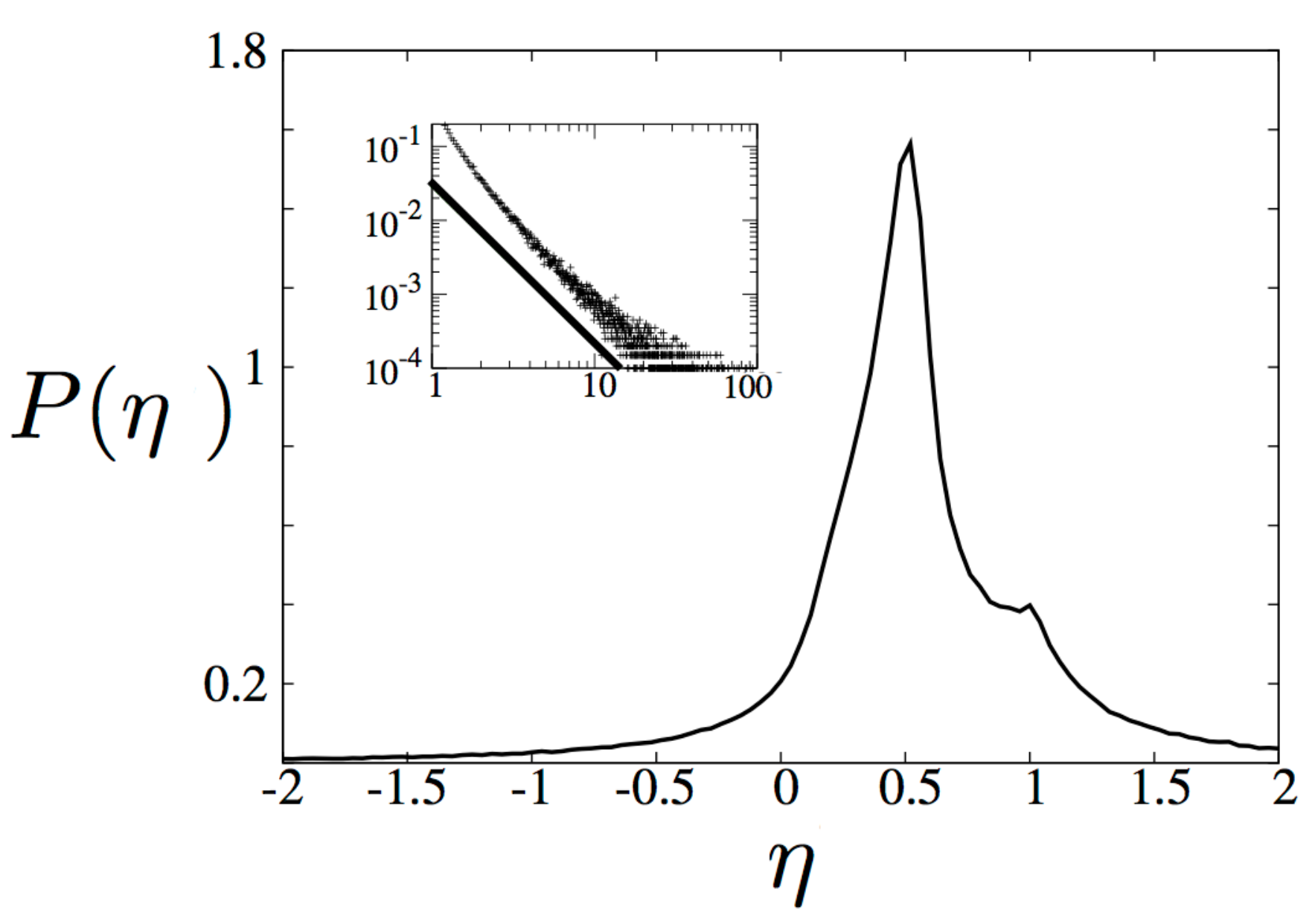}
\caption{Probability distribution of $\eta$ at $\tau=0.5$ at $T_h=1$. Here $\la\eta\ra=0.54$ and $\sigma_{\eta}=3.56$. Inset: the positive tail of the distribution  behaves as a power law with exponent -2.155 $\pm$ 0.014 ( slope of the straight line).}
\label{oeffdist}
\end{center}
\end{figure}

We have plotted the distribution of stochastic efficiency in Fig. \ref{oeffdist} at $\tau=0.5$ and $T_h=1.0$, a typical point of the phase diagram where our system works as stochastic heat engine. We found that only for $40.3\%$ cycles, the system works as a heat engine. Thus, even in overdamped regime, the engine operation is unreliable. Both $P(\epsilon)$ and $P(\eta)$ show power-law decays in their tails for several decades. The exponent values are given in the figure captions. The exponent of $P(\epsilon)$ is different from that we have obtained in case of underdamped dynamics.  


\section{Conclusion}

We will conclude by focusing on the main  results to point out the crucial differences between macro and micro refrigerators. In the macro-world, idealized refrigerator is like an engine running backward in time, i.e., using the work to extract heat from the cold bath and dumping the heat at hot bath. We have shown that such a simple picture is not valid for single particle heat engine at nano-scale. The engine running in a reverse order may produce refrigerator, stochastic heater of type I and II as well as unreliable heat engine. For our model of study, fluctuations in stochastic COP and efficiency  dominate their mean values even in the quasistatic case. This implies that in such a situation mean is not a good physical variable and one must study the behaviour of full probability distributions which in all our cases contain power law behaviour in their tails with varied exponents. Stochastic COP and efficiency may exceed Carnot bound. However, averaged COP and efficiency, as defined here, are bounded by the Carnot value. This can readily be shown by fluctuation theorem of heat engines.

 In every cycle, a macro thermal machine under refrigerating protocol acts as a refrigerator. This is not true for micro refrigerators. We know that the trajectory dependent work, heat and COP are randomly distributed variables over the cycles of a micro refrigerator. The cycle time can be large or small, but there will always be a considerable number of cycles where the system will not run as a refrigerator due to fluctuations of these thermodynamic variables. This makes the micro refrigerator unreliable in comparison to macro refrigerators. This reliability, quantified as a fraction of cycles running as a refrigerator, increases as we increase the cycle time $\tau$. The obtained heat engine is most unreliable.   

 The phase diagram and  thermodynamic quantities describing the micro refrigerators as well as the micro heat engines are crucially protocol dependent. We have noticed in our other studies (will be published elsewhere), that the micro refrigerators / heat engines, running under Stirling protocol (which is very similar to Carnot protocol but devoid of the adiabatic jumps), though produce qualitatively similar phase diagrams as Carnot-protocol, are less fluctuating in large cycle time limit. Moreover, only in the 
underdamped case in a small parameter space heat engine operation is possible. The adiabatic jumps are key players behind the unreliability of micro heat engines in \cite{rana14} or refrigerators described in earlier sections. Only in the overdamped quasistatic limit we have shown that for our given specific protocol, refrigerator can work in a reversible mode of operation provided $\frac{T_h}{T_L}=2$. This clear from zero entropy production for this case (e.g., see Fig. \ref{ow} D). Power output is found to be an useful optimization criterion for finite time heat engines. On the other hand, many optimization criteria were proposed in case of refrigerator. Widely used target function was introduced in \cite{yan90,vel97,tom12,tom13,yuan14} by giving equal footing for COP and cooling rate, known as $\chi$-criterion \cite{tom12,tom13}. In view of new findings in Brownian refrigerators, it will be of immense  interest to study these quantities in our system. It may be noted that our studied engine or refrigerator do not reach Carnot limit in the efficiency / COP in the quasistatic regime. The engine is not a microscopic version of Carnot engine. There are other engines which reproduces Carnot limit for efficiency  in the quasistatic regime \cite{mar14}. Studies on differences in these two types of engines will be useful. Work along this direction is in progress.

\section{ACKNOWLEDGMENTS}
Authors A.S., S.R. and P.S.P thanks  Edgar Roldan for several useful discussions. A.M.J. thanks Department of Science and Technology, India for financial support.

\end{document}